\documentstyle[12pt,psfig,epsf]{article}
\newcommand{\be}{\begin{equation}}
\newcommand{\qee}{\end{equation}}
\begin{document}
\begin{titlepage}
\title{
{\bf Gonihedric String Equation I}
}
{\bf
\author{ 
G.K.Savvidy\\
National Research Center "Demokritos",\\
Ag. Paraskevi, GR-15310 Athens, Greece \\
\vspace{1cm}\\
K.G.Savvidy\\
Princeton University, Department of Physics\\
Princeton, New Jersey 08544-0708, USA
}
}
\date{}
\maketitle
\begin{abstract}
\noindent

We discuss the basic properties of the gonihedric string 
and the problem of its 
formulation in continuum. We propose a generalization of the  
Dirac  equation and of the corresponding gamma matrices
in order to describe the gonihedric string.
The wave function and the Dirac matrices are infinite-dimensional. 
The spectrum of the theory consists of 
particles and antiparticles of increasing half-integer spin 
lying on quasilinear trajectories of different slope. Explicit formulas for 
the mass spectrum allow to compute the string tension and thus demonstrate 
the string character of the theory. The string tension varies from trajectory to 
trajectory and indicates nonperturbative character of the spectrum.
Additional $\Gamma_{5}$ and pure Casimir mass terms in the string equation 
allow to increase the slope of the trajectories, so that the mass spectrum grows
as $j^{3}$ and $j^{5}$. The equation does not admit tachyonic solutions, 
but still has unwanted ghost solutions. We include bosons and show that
they also lie on the same quasilinear trajectories.

\end{abstract}
\thispagestyle{empty}
\end{titlepage}
\pagestyle{empty}

\section{Introduction}
\vspace{.5cm}

There is some experimental and theoretical evidence for the existence of a
string theory in four dimensions which may describe strong interactions and 
represent the solution of QCD \cite{nilsen}. 

One of the possible candidates for that
purpose is the gonihedric string which has been defined as a model of 
random surfaces with an action which is proportional to the linear size 
of the surface \cite{sav2}

\be
A(M_{2}) = m\sum_{<ij>} \lambda_{ij}
\cdot \Theta(\alpha_{ij}),~~~~~~~ \Theta(\alpha)= \vert 
\pi - \alpha \vert^{\varsigma} , \label{action}
\qee
where $\lambda_{ij}$ is the length of the edge $<ij>$ of the 
triangulated surface $M_{2}$ and $\alpha_{ij}$ is the dihedral angle 
between two neighbouring triangles of $M$ sharing a common edge $<ij>$.
The angular factor $\Theta$ defines the rigidity of the random surfaces 
\cite{sav2}. The action has been defined for self-intersecting surfaces 
as well \cite{sav2} and is equal to 

\be
A(M_{2}) = m\sum_{<ij>} \lambda_{ij}
\cdot \Theta(\alpha_{ij})~~ + ~~k_{r}~m\sum_{<ij>} \lambda_{ij}
\cdot (\Theta(\alpha^{(1)}_{ij})+...+\Theta(\alpha^{(r)}_{ij}))
\qee
where the number of terms inside parentheses is equal to the order of the 
intersection $r=3,4,...$;~~$r$ is the number of triangles sharing the 
edge $<ij>$.
The coupling constant $k_{r}$ is called self-intersection coupling constant 
\cite{sav3}
\footnote{Both terms in the action have the same 
dimension and the same geometrical nature: the action (\ref{action})  
"measures"  two-dimensional surfaces in terms of length, while self-intersections,
one-dimensional manifolds to start with, are also measured 
in terms of length.}.

The model has a number of properties which make it very close to the Feynman
path integral for a point-like relativistic particle. This can be seen from 
(\ref{action}) in the limit when the surface $M_{2}$ degenerates into a single 
world line $M_{1}$ (see Figure 9), in that case 

\be
A(M_{2})~~ =~m\int_{M_{2}}(\frac{1}{R_1} + 
\frac{1}{R_2})~dS~~ \rightarrow ~2\pi 
m\int_{M_{1}}~dL~~=~~L(M_{1}).               \label{basicpro}
\qee
In this limit the classical equation of motion for the gonihedric string

$$K~=~\frac{1}{R_{1}~R_{2}}~ =~0,$$
which describes the evolution of rigid string, is reduced to the classical 
equation of motion for a free relativistic particle\footnote{For simplicity 
we present the classical equation of motion in three dimensions.}.

The other important property of the theory is that 
at the classical level the string tension is equal to zero and 
quarks viewed as open 
ends of the surface are propagating freely without interaction  \cite{sav2}:

$$\sigma_{classical}=0 .$$
This is because the gonihedric action (\ref{action}) 
is equal to the perimeter $P(M_{2})$ of the flat Wilson loop 

$$
A(M_{2}) \rightarrow ~~~~P(M_{2})=R+T
$$
and the potential $V(R)$ is constant.
As it was demonstrated in \cite{sav2}, quantum fluctuations generate the 
nonzero string tension 
\be
\sigma_{quantum}= \frac{d}{a^2}~(1 - ln \frac{d}{\beta}) , \label{tension}
\qee
where $d$ is the dimension of the spacetime, $\beta$ is the coupling constant,
$a$ is the scaling parameter and $\varsigma = (d-2)/d$
in (\ref{action}).
In the scaling limit $\beta \rightarrow \beta_{c}=d/e$ the string tension 
has a finite limit while the scaling parameter tends to zero as
$a=(\beta - \beta_{c})^{1/2}, $
thus the critical exponent $\nu$ is equal to one half, 
$\nu = 1/d_{H} = 1/2$ , where $d_{H}$ is the Hausdorff dimension.

Thus at the tree level the theory describes free quarks with the string
tension equal to zero, quantum fluctuations generate nonzero string 
tension and, as a result, the quark confinement \cite{sav2}. 
The gonihedric string may 
consistently describe asymptotic freedom and confinement as it is expected
to be the case in QCD and we have to ask what type of 
equation may describe this string. The aim of this article is to  
answer this question.

\vspace{1.0cm}
Some additional understanding of the physical behaviour of the system comes
from the transfer matrix approach \cite{sav4}. The transfer matrix can be 
constructed in two cases, $k=0$ and $k= \infty$, that is for 
$free-intersecting$ and $self-avoiding~surfaces$. In both cases it describes the 
propagation of the closed string $M^{\tau}_{1}$ in time direction $\tau$ with an 
amplitude which is proportional to the sum of the length of the string  
and of the total curvature \cite{sav4}
$$ 
A(M^{\tau}_{1}) = \int_{M^{\tau}_{1}} dL    +   \int_{M^{\tau}_{1}}k(L)dL
$$
and of the interaction which is proportional to the overlapping length 
of the string on two neighbouring time slices
$$ 
A_{int}(M^{\tau};M^{\tau +1}) = \int_{M^{\tau} \cap \ M^{\tau +1}} dL .   
$$
Considering the system 
of closed paths on a given time slice $\tau$ as a separate system with the 
length and curvature amplitude one can see that   
this two-dimensional system has a 
continuum limit which can be described by a 
free Dirac fermion \cite{sav4}.  
Thus on every  time slice the system generates a fermionic string which may 
propagate in time because of the week interaction between neighbouring 
time slices \cite{sav4}. 
Thus the physical picture of the fermionic string propagation which follows
from the transfer matrix approach again stresses the need for answering the 
question of the corresponding string equation.

\vspace{1.cm}
In addition to the formulation of the theory in the continuum space 
the system allows an equivalent representation on Euclidean lattices 
where a surface is associated with a collection of plaquettes 
\cite{wegner,sav3}.
Lattice spin systems whose interface energy coincides with the action 
(\ref{action}) have been constructed in an arbitrary dimension $d$ \cite{wegner} 
for the self-intersection coupling constant $k=1$ and for an arbitrary 
$k$ in \cite{sav3}. This gives an opportunity for numerical simulations 
of the corresponding statistical systems in a way which is similar to the 
Monte Carlo simulations of QCD \cite{creutz}. The Monte Carlo 
simulations \cite{cut} demonstrate that the gonihedric 
system with a large intersection coupling constant undergoes the 
second order phase transition and the string tension is generated 
by quantum fluctuations, as it was  expected theoretically \cite{sav2}.
This result again suggests the existence of 
a noncritical string theory in four dimensions.

\vspace{1.0cm}
It is natural to think that each
particle in this theory should be viewed as a state of a 
complex fermionic system and that this system should have a point-particle 
limit when there is no excitation of the internal motion, thus requiring the 
basic property of the gonihedric string (\ref{basicpro}).
The question is then how to incorporate this internal motion into existing 
point particle equations. Ettore Majorana suggested in 1932 \cite{majorana} an 
extension of the Dirac equation by constructing an infinite-dimensional 
representation
of the Lorentz group and the corresponding extension of the gamma matrices. 
The equation has the Dirac form
\be
\{~i~\Gamma_{\mu}~\partial^{\mu}~~-~~M~\}~~\Psi~~~=0  \label{stringeq1}
\qee
and the Majorana commutation relations which define the $\Gamma_{\mu}$ matrices
are given by the formula (see (13) in \cite{majorana})
\be
[\Gamma_{\mu},~I_{\lambda \rho}] = \eta_{\mu \lambda}~\Gamma_{\rho}
- \eta_{\mu \rho}~\Gamma_{\lambda},
\qee
where $I_{\mu \nu}$ are the generators of the Lorentz algebra. These equations 
allow to find the $\Gamma_{\mu}$ matrices when the representation of the 
$I_{\mu \nu}$ is given. The original Majorana solution 
for $\Gamma_{\mu}$ matrices is infinite-dimensional (see equation (14) in 
\cite{majorana}) and the mass spectrum of the theory is equal to
\be
M_{j}=\frac{M}{j+1/2},               \label{majorana}
\qee
where $j=1/2,3/2,5/2,....$ in the fermion case and $j=0,1,2,....$ in the 
boson case.
The main problems in the Majorana theory are the decreasing mass spectrum 
(\ref{majorana}), 
absence of antiparticles and troublesome tachyonic solutions - the problems 
common to high spin theories. Nevertheless we intend to 
interpret the Majorana theory as a natural way to incorporate
$additional$ degrees of freedom into the relativistic Dirac equation.   
The problem is to formulate physical principles allowing to choose 
appropriate 
representations of the Lorentz group in order to have a string 
equation with necessary 
properties. The above discussion of the gonihedric string shows  
that an appropriate equation should exist. Indeed 
we  shall demonstrate that the  solution of the 
Majorana commutation relations exists and the corresponding 
equation has an increasing mass spectrum 
(\ref{dmassspec}), (\ref{dmassspec5}) and 
nonzero string tension (\ref{strtension}).

\vspace{1.0cm}
An alternative way to incorporate the internal motion into the Dirac equation 
was suggested by Pierre Ramond in 1971 \cite{ramond}. 
In his extension of the Dirac equation the internal motion 
is incorporated through the construction of operator-valued gamma matrices. 
The equations which define the $\Gamma_{\mu}$ matrices are
$$
<\Gamma_{\mu}(\tau)> = \gamma_{\mu},
$$
$$
\{ \Gamma_{\mu}(\tau), \Gamma_{\nu}(\tau^{'}) \} = 
2 \eta_{\mu \nu} ~\delta(~ \frac{1}{2\pi\alpha^{'} } ~(\tau -\tau^{'})~ )
$$
$$
\Gamma^{+}_{\mu}(\tau)~\gamma_{0}=
\gamma_{0}~\Gamma_{\mu}(\tau),
$$
where it is required that the proper-time average $<...>$ 
over the periodic internal motion with period 
$2\pi \alpha^{'} =1 /\sigma $ should coincide with the Dirac 
matrices. The mass spectrum lies on the linear trajectories 
$$
M^{2}_{n} = \frac{1}{\alpha^{'}}~(j + n -3/2)~~~~~~~~~~~~~n=1,2,3,....
$$
where $\sigma$ is the string tension, $j=1/2,3/2,...$ and $n$ 
enumerates the trajectories (here part of the states are spurious). 
In both cases one can see effective extensions of Dirac gamma
matrices into the infinite-dimensional case, but these extensions are 
quite different. The free Ramond string is a consistent theory 
in ten dimensions 
and the spectrum contains a massless ground state
\footnote{For the subsequent development of superstring theories and their  
unification into a single M-theory see \cite{15years}}.

\vspace{1.0cm}
For our purposes we shall follow Majorana's approach to incorporate the 
internal motion in the form of an infinite-dimensional wave equation.
Unlike Majorana we shall consider the {\it infinite sequence} of high-
dimensional representations of the Lorentz group with  nonzero Casimir
operators $(\vec{a}\cdot\vec{b})$ and $(\vec{a}^2 -\vec{b}^2)$. 
The important restriction which we have to impose on the 
system is that it should have a point particle limit. In the given case 
this restriction should be 
understood as a principle according to which  the infinite sequence of 
representations should contain the Dirac spin one-half representation.   
In the next section we shall review the known representations 
of the Lorentz group $SO(3,1)$ and shall select necessary 
representations for our purpose. These representations $(j_{0},\lambda)$ and 
their adjoint $(j_{0},-\lambda)$ are enumerated by the 
index $r=j_{0} +1/2$, where $r=1,...,N$ and $j_{0}=1/2,3/2,...$~ is 
the lower spin in the representation $(j_{0},\lambda)$, thus 
$j=j_{0},j_{0}+1,...$. 
The representations used in the Dirac equation
are $(1/2,-3/2)$ and $(1/2,3/2)$ and
in the Majorana equation they are $(0,1/2)$ in the boson case and $(1/2,0)$ in
the fermion case. 
We shall introduce also the concept of the {\it dual representation} which is 
defined as 
$\Theta = (j_{0};\lambda)  \rightarrow (\lambda;j_{0})
= \Theta^{dual}$. This dual transformation is essentially used in subsequent 
sections to construct the 
solution of the Majorana commutation relations which has an increasing 
mass spectrum and is bounded from bellow.

The invariant Lagrangian, the string equation  and the Majorana commutation 
relations are formulated in the third and  forth sections. 
In the fifth and six sections 
the solution of Majorana commutators is constructed 
for small values $N=1,2,....$ in the form of Jacoby matrices \cite{minnes}
and then it is used to construct the solution 
for an arbitrary $N$. 
This basic solution (we shall call it $B$-solution) is well defined for any $N$ 
and allows to take 
the limit $N \rightarrow \infty$. The spectrum can be computed for any $N$ 
and demonstrates a tendency to concentrate around the eigenvalues
$+1$ and $-1$. When $N \rightarrow \infty$ all masses are equal to one and  
the spectrum is bounded from bellow, but 
the $\Gamma_{0}$ matrix is not Hermitian. In the next four sections we 
reach the necessary Hermitian property of the system and the correct 
commutation relation
with the matrix of the invariant form. The solution is symmetric and we shall 
call it $\Sigma$-solution.
The spectrum of the theory now 
consists of particles and antiparticles of increasing half-integer spin and 
lying on quasilinear trajectories of different slope
\be
2\pi\sigma_{n}  = \frac{1}{\alpha^{'}_{n}} = \frac{2M^2}{n}~~~~~~n=1,2,3....
\label{strtension}
\qee
here $n$ enumerates the trajectories and the lower spin on a 
trajectory is $j=n+1/2$.
{\it This result demonstrates that we have indeed a string equation which has 
trajectories with different string tension and that trajectories with large $n$
are almost "free" because the string tension tends to zero.}
The number of particles with a given spin $j$ is equal to $j+1/2$. 
The components of the wave function which 
describe spin $j$ satisfy a partial differential equation of order $2j+1$. The 
corresponding differential operator is a sum of the powers of the 
D'Alembertian. The general formula for all trajectories is
$$
M^{2}_{n}= \frac{2 M^{2}}{n} ~\frac{j^2 -(2n-1)j +n(n-1)}{j-(n-1)/2}~~n=1,2,...
$$
where $j = n+1/2, n+5/2, ....$. The unwanted property of the $\Sigma$-solution 
is that the smallest mass   
on a given trajectory $n$ has the spin $j = n+1/2$ and decreases as 
$M^{2}_{n}(j=n+1/2)= 3M^{2}/n(n+3)$,~~~the disadvantage 
which is common with the Majorana solution (\ref{majorana}). 
In the eleventh section we introduce additional 
$(\vec{a}\cdot\vec{b})~\Gamma_{5}$ and pure Casimir $(\vec{a}^2 -\vec{b}^2)$ 
mass terms into the string equation (\ref{stringeq1})
\be
\{~i~\Gamma_{\mu}~\partial^{\mu}~~-~~M~(\vec{a}
\cdot\vec{b})\Gamma_{5}~~-~~gM~(\vec{a}^2 -\vec{b}^2)~\}~~\Psi~~~=0  
\qee
where $(\vec{a}\cdot\vec{b})$ and $(\vec{a}^2 -\vec{b}^2)$ are the Casimir 
operators of the Lorentz algebra. 
These terms essentially increase the  string tension so that all trajectories 
have a nonzero slope, but still we have a decreasing part in the spectrum. 

The cardinal solution of the problem is given in the 
twelfth section where we perform a $dual$ transformation of the system. 
The dual formula for the mass spectrum is
\be
M^{2}_{n}= \frac{2M^2}{n}~\frac{(j+n)(j+n+1)}{j+(n+1)/2},~~~~~~~~~~~~n=1,2,...
\label{dmassspec}
\qee
The lower spin on a given 
trajectory is either $1/2$ or $3/2$ depending on n: if n is odd then 
$j_{min}=1/2$, if n is even then $j_{min}=3/2$. The essentially new property
of the dual equation is that  now we have an infinite number of states with 
a given spin $j$ instead of $j+1/2$, which we had before we did 
the dual transformation.
String tension has the same values as in (\ref{strtension})
and the lower mass on a given trajectory $n$ is given by the formula 
$M^{2}_{n}(j=1/2) =4M^{2}(2n+1)(2n+3)/n(n+2) \rightarrow (4M)^2$
and the similar one for $M^{2}_{n}(j=3/2)$.
{\it 
Thus the main problem of decreasing spectrum 
has been solved after the dual transformation because the 
spectrum is now bounded 
from below.} Additional $\Gamma_{5}$ mass term is also analyzed with 
the main result
\be
M^{2}_{n}= \frac{2M^2}{n}\frac{(j+n)^{2}(j+n+1)^{2}}{j+(n+1)/2} 
\frac{1}{4}~~~~~n=1,2,...    \label{dmassspec5}
\qee
and the pure Casimir mass term with the following spectrum 
\be
M^{2}_{n}= g^{2}~\frac{2M^2}{n}\frac{(j+n)^{3}(j+n+1)^{3}}
{j+(n+1)/2}~~~~~n=1,2,3,...    \label{casimisspec}
\qee
In the last thirteenth section we extend the equation to include bosons 
and show that bosons also lie on the same quasilinear trajectories 
(\ref{dmassspec}). At the end we discuss the problems of tachyonic and 
ghost solutions of the Majorana and of the new  equation. 
Some technical details can be found in the Appendixes.

\section{Representations of Lorentz Algebra}

The algebra of the $SO(3,1)$ generators \cite{heisenberg,majorana}
\begin{eqnarray}
[I^{\mu \nu} ~ I^{\lambda \rho}] = -\eta^{\mu \lambda}~I^{\nu \rho} +
\eta^{\nu \lambda}~I^{\mu \rho} + \eta^{\mu \rho}~I^{\nu \lambda} - 
\eta^{\nu \rho}~I^{\mu \lambda}\nonumber
\end{eqnarray}
can be rewritten in terms of $SO(3)$ generators $\vec{a}$ and 
Lorentz boosts $\vec{b}$
\begin{eqnarray}
a_{x}  = iI^{23} ~~~~ a_{y} = iI^{31}~~~~a_{z} = iI^{12}\nonumber\\
b_{x}  = iI^{10} ~~~~ b_{y}=  iI^{20}~~~~b_{z} = iI^{30}\nonumber
\end{eqnarray}
as \cite{majorana} (we use Majorana's notations)
\be
[a_{x},  a_{y}]= ia_{z} \label{alge1}
\qee
\be
[a_{x},  b_{y}]= ib_{z} \label{alge2}
\qee
\be
[b_{x},  b_{y}]= -ia_{z} . \label{alge3} 
\qee
The irreducible representations $R^{(j)}$ of $SO(3)$ algebra (\ref{alge1}) are 
\begin{eqnarray}
<j,m\vert ~a_z ~\vert j,m> =m \nonumber\\
<j,m\vert ~a_{+} ~\vert j,m-1> = \sqrt{(j+m)(j-m+1)} \nonumber\\
<j,m\vert ~ a_{-} ~\vert j,m+1> = \sqrt{(j+m+1)(j-m)}, \label{matr}
\end{eqnarray}
where $m=-j,...,+j$,~~~the dimension of  $R^{(j)}$ is~~ $2j+1$~~ and 
$j =0,~ 1/2,~1,~3/2,~...$ Then
$$
<j,m\vert ~\vec a^{2} ~\vert j,m> = j(j+1).\nonumber
$$
Since $\vec{b}$ transforms as a vector under $SO(3)$ spatial 
rotations (\ref{alge2}) it follows that
$$
<j \vert~\vec{b}~\vert j'> =0\nonumber
$$
unless $j'=j$, or $j'=j~\pm 1$, thus the matrix elements of $\vec{b}$
are fixed by their vector character, apart from the factors 
depending on $j$ but not on $m$. Therefore 
the solution of the commutation relations (\ref{alge2}) 
for $\vec{b}$ can be parameterized as \cite{heisenberg,majorana}
\begin{eqnarray}
<j,m \vert ~b_{z} ~\vert j,m> = \lambda_{j} \cdot m \label{7}\nonumber \\
<j-1,m \vert~ b_{z}~ \vert j,m> =\varsigma_{j}\cdot \sqrt{(j^2-m^2)} \nonumber\\
<j,m\vert~ b_{z}~ \vert j-1,m> = \varsigma_{j} \cdot \sqrt{(j^2-m^2)}\nonumber
\end{eqnarray}

\begin{eqnarray}
<j,m \vert ~b_{+}~ \vert j,m-1> =~ \lambda_{j} \cdot \sqrt{(j+m)(j-m+1)}
\nonumber\\
<j-1,m \vert ~b_{+}~ \vert j,m-1> =~  \varsigma_{j} \cdot \sqrt{(j-m)(j-m+1)}
\nonumber\\
<j,m \vert~ b_{+}~ \vert j-1,m-1> = - \varsigma_{j} \cdot 
\sqrt{(j+m)(j+m-1)}\nonumber
\end{eqnarray}

\begin{eqnarray}
<j,m \vert ~b_{-} ~\vert j,m+1> =~ \lambda_{j} \cdot \sqrt{(j-m)(j+m+1)}
\nonumber\\
<j-1,m\vert ~b_{-}~ \vert j,m+1> = - \varsigma_{j}\cdot \sqrt{(j+m)(j+m+1)}
\nonumber\\
<j,m\vert ~b_{-}~ \vert j-1,m+1> =~ \varsigma_{j} \cdot \sqrt{(j-m)(j-m-1)}
                                                                   \label{19}
\end{eqnarray}
where the amplitudes $\lambda_{j}$ describe {\it diagonal} transitions 
inside the $SO(3)$ multiplet $R^{(j)}$ while  $\varsigma_{j}$ describe
{\it nondiagonal} transitions between $SO(3)$ multiplets which form the 
representation $\Theta$ of $SO(3,1)$. 
These amplitudes should satisfy the boundary condition
\be
(j_0 + 1)\lambda_{j_{0}}= i~\lambda~~~~~~~\varsigma_{j_{0}}=0 ,                                 \label{boun}
\qee
where $j_{0}$ defines the lower spin in the representation
$\Theta=(j_{0};\lambda)$ and $\lambda$ is a free parameter, thus 
\be
\Theta(j_{0},\lambda) = \oplus \sum^{\infty}_{j=j_{0}}~R^{(j)} . \label{repres}
\qee
The amplitudes $\lambda$ and $\varsigma$ can 
be found from the last of the commutation relations (\ref{alge3}) rewritten in 
the component form  
\begin{eqnarray}
(2j+3) \cdot \varsigma^{2}_{j+1} - 
(2j-1)\cdot \varsigma^{2}_{j} - \lambda^{2}_{j} =1 
\nonumber\\
\{(j+1) \cdot \lambda_{j} - (j-1) \cdot \lambda_{j-1} \} \cdot \varsigma_{j} 
= 0 . \label{etav}
\end{eqnarray}
From the last recursion equation and the boundary conditions 
(\ref{boun}) it follows that  
\be
\lambda_{j}~ = ~\frac{j-1}{j+1}~ \lambda_{j-1} 
=~~ \frac{j_{0}(j_{0}+1)}{j(j+1)}~\lambda_{j_{0}} 
=~~ i\frac{j_{0}~\lambda}{j(j+1)},                          \label{lamb}
\qee
where $j_{0}$ is the lower spin in the representation
$\Theta = (j_{0};\lambda)$ and 
$\lambda$ appears as an essential dynamical parameter which 
cannot be determined solely from the kinematics of the Lorentz group.
Substituting $\lambda_j$ from (\ref{lamb}) into (\ref{etav})
one can find $\varsigma_j$
\be
\varsigma^{2}_{j} = \frac{(~j^{2} -j^{2}_{0}~)~(~j^2 -\lambda^{2})}
{j^2~(~4j^2-1~)}         \label{nond}
\qee
in terms of $j_{0}$ and $\lambda$. The adjoint representation is defined as
$\dot{\Theta} = (j_{0};-\lambda)$ and we shall 
introduce here also the concept of the {\it dual representation} 
which we define as

\be
\Theta = (j_{0};\lambda)  \longleftrightarrow (\lambda;j_{0})
= \Theta^{dual} \label{dual}
\qee 
The representations used in the Dirac equation are $(1/2,-3/2)$ and $(1/2,3/2)$
and
in the Majorana equation they are $(0,1/2)$ in the boson case and $(1/2,0)$ in
the fermion case. The infinite-dimensional Majorana representation $(1/2,0)$ 
contains $j=1/2,3/2,...$ 
multiplets of the $SO(3)$ while $(0,1/2)$ contains $j=0,1,2,...$ multiplets.
The essential difference between these representations is 
that the Lorentz boost operators $\vec{b}$ are  
diagonal in the first case ( the diagonal amplitudes (\ref{lamb}) are 
$\lambda^{1}_{1/2} = -
\lambda^{\dot{1}}_{1/2}= i$ and nondiagonal amplitudes (\ref{nond}) are 
equal to zero
$\varsigma^{1}_{3/2}=\varsigma^{\dot{1}}_{3/2}=0$)  and they are nondiagonal
in the case of Majorana
representations (the nondiagonal amplitudes are $\varsigma_{j}=1/2$ and 
the diagonal amplitudes are equal to zero $\lambda_{j}=0$ as it follows 
from  (\ref{lamb}) and (\ref{nond}) and coincide  with the 
solution (19) in \cite{majorana} ).

\vspace{1cm}
Let us consider pairs of adjoint representations
with $j_{0}=1/2,~3/2,~\cdots$
\be
\Theta_{r} = (j_{0};  \lambda)~~~~~~~~~\Theta_{\dot{r}} = 
(j_{0};- \lambda)       \label{ourrep}
\qee
which we shall enumerate by the representation
index $r=1,2,3,...$~~~ so that $j_{0}=r-1/2$ (see Fig.1).
The corresponding matrices $\vec{b}^{r}$ and $\vec{b}^{\dot{r}}$ are 
defined by (\ref{19}) where from (\ref{lamb}) and 
(\ref{nond}) we have
\be
\lambda^{r}_{j} = - \lambda^{\dot{r}}_{j} 
= i ~\frac{r -1/2}{j(j+1)} ~\lambda ~~~~~~~~j\geq r-1/2
\qee
and 
\be
\varsigma^{r}_{j} = \varsigma^{\dot{r}}_{j}= 
\frac{1}{2} \sqrt{(1-\frac{r^2-r}{j^2-1/4})(1-(\frac{ 
\lambda}{j})^2)}~~~~~~j \geq r+1/2
\qee
and we shall consider the case when 
\be
-3/2 \leq \lambda \leq 3/2  \label{ourlambda}
\qee
to have $\varsigma$ real for all values of $r$.
The Casimir operators $(\vec{a}\cdot \vec{b})$ and $~(\vec{a}^2 - 
\vec{b}^2)~$
for the representation $\Theta_{r}$ are equal 
correspondingly  to 
\begin{eqnarray}
<j,m\vert ~\vec{a} \cdot \vec{b}~ \vert j,m> = i~\lambda~ (r -1/2)\\
<j,m\vert ~(\vec{a}^2 -\vec{b}^2)~ \vert j,m> = (r-1/2)^2 + 
\lambda^2 -1  .   
\end{eqnarray}
As it is easy to see from these formulas the Casimir operator 
$(\vec{a}\cdot \vec{b})$ is nonzero only if $\lambda \neq 0$. For the 
Majorana representations $(1/2,0)$  and $(0,1/2)$ the Casimir operator 
$(\vec{a}\cdot \vec{b})$ is equal to zero.

\section{The Invariant form, Lagrangian and conserved current}

To have an invariant scalar product 
\be
<~\Theta ~\Psi_1~\vert~ \Theta ~\Psi_2~>~~~ =~~~<~\Psi_1~\vert~ \Psi_2~> 
\qee
where $\Theta = 1 + \frac{1}{2}\epsilon_{\mu \nu}~I^{\mu \nu}$, 
one should define the matrix $\Omega$  
\be
<~\Psi_1~\vert~ \Psi_2~> = \bar{\Psi}_{1}~\Psi_{2} =\Psi^{+}_{1}~\Omega~\Psi_{2}
 = \Psi^{*~r}_{1~jm}~\Omega^{rr'}_{jm~j'm'}~\Psi^{r'}_{2~j'm'}
\qee
with the properties
\begin{eqnarray}
\Omega~a_{k} = a_{k}~\Omega\\
\Omega~b_{k} = b^{+}_{k}~\Omega  \label{gode} \\
\Omega = \Omega^{+}.
\end{eqnarray}
From the first relation it follows that 
\be
\Omega = \omega^{rr'}_{j}\cdot\delta_{jj'}\cdot\delta_{mm'}
\qee
and from the last two equations, for our choice of the representation $\Theta$
(\ref{ourrep})-(\ref{ourlambda}) and for a real $\lambda$, 
\be
\omega^{r\dot{r}}_{j}=\omega^{\dot{r}r}_{j}~=~1~~~\omega^{2}_{j}=1,
\qee
thus $\Omega$ is an antidiagonal matrix
\be
\bar{\Psi}_{1}~\Psi_{2}~~ =~~\sum^{\infty}_{r=1}~\sum^{\infty}_{j=r-1/2} 
\sum^{j}_{m=-j}~(~\Psi^{*~r}_{1~jm}~\Psi^{~\dot{r}}_{2~jm} + 
\Psi^{*~\dot{r}}_{1~jm}~\Psi^{~r}_{2~jm}~)     \label{form}
\qee
and to have a finite invariant product the summation should be convergent
\footnote{The modulus of the wave function is given by the formula
$$
\Psi^{+}_{1}~\Psi_{2}~~ =~~\sum^{\infty}_{r=1}\sum^{\infty}_{j=r-1/2} 
\sum^{j}_{m=-j}~(~\Psi^{*~r}_{1~jm}~\Psi^{~r}_{2~jm} + 
\Psi^{*~\dot{r}}_{1~jm}~\Psi^{~\dot{r}}_{2~jm}~).
$$
}.

Having in hand the invariant form (\ref{form}) one can construct the 
Lagrangian  
\be
L = \int~\{i~\bar{\Psi}~\Gamma_{\mu}~\partial^{\mu}~
\Psi~-i~\partial^{\mu}~\bar{\Psi}~\Omega \Gamma^{+}_
{\mu} \Omega~\Psi~-2~M~\bar{\Psi}~\Psi~\}~d^{4}x
\qee
and the corresponding equation of motion 
\be
\{~i~(\Gamma_{\mu}+\Omega \Gamma^{+}_{\mu} 
\Omega )~\partial^{\mu}~~-~2~M~\}~~\Psi~~~=0 .  
\label{eqmo}
\qee
Multiplying the conjugate equation 
\be
-i~\partial^{\mu}~\Psi^{+}~(\Omega \Gamma_{\mu}\Omega + 
\Gamma^{+}_{\mu})~-2M~\Psi^{+}~~=0
\qee
by $\Omega$ from the r.h.s. we have
\be
-i~\partial^{\mu}~\bar{\Psi}~(\Gamma_{\mu}+\Omega \Gamma^{+}_{\mu} 
\Omega)~-2M~\bar{\Psi}~~=0
\qee
and then from both equations follows the conservation of the current 
\be
J_{\mu} = i~\bar{\Psi}~\frac{(\Gamma_{\mu}+
\Omega \Gamma^{+}_{\mu} \Omega)}{2}~\Psi~~~~~~~~~\partial^{\mu}~J_{\mu}=0.
\qee
The crucial point is that current density should be  positive definite
\be
J_{0} = i~\Psi^{+}~\frac{(\Omega \Gamma_{0}+\Gamma^{+}_{0} \Omega)}{2}~\Psi,
\qee
which is equivalent to the positivity of the eigenvalues of the matrix
\be
\rho = \frac{\Omega~\Gamma_{0}+\Gamma^{+}_{0}~\Omega}{2}.
\qee
If in addition the important relation is required to be satisfied 
\be
\Gamma^{+}_{\mu}~\Omega =\Omega~\Gamma_{\mu}, \label{holdc}
\qee
then the current will have the form 
\be
J_{\mu} =  i~\bar{\Psi}~ \Gamma_{\mu}~\Psi, 
\qee
and equation (\ref{eqmo}) can be transformed in this case 
into Hamiltonian form
\be
i~\frac{\partial~\Psi}{\partial~t} = H~\Psi
\qee
with the Hamiltonian
\be
H = \Gamma^{-1}_{0}~\vec{\Gamma}~\vec{p}~+~\Gamma^{-1}_{0}~M .
\qee

\section{Majorana commutation relations for Gamma matrices}

The Lagrangian and the equation  
\be
\{~i~\Gamma_{\mu}~\partial^{\mu}~~-~~M~\}~~\Psi~~~=0  \label{stringeq}
\qee
should be invariant under Lorentz transformations
\be
X'_{\mu} =\Lambda_{\mu}^{~\nu}~ X_{\nu}~~~~~~~~~~\Psi'(X')=
\Theta(\Lambda)~ \Psi(X),
\qee
which  leads to the following equation for the gamma matrices 
\be
\Gamma_{\nu} = \Lambda_{\nu}^{~\mu}~\Theta~\Gamma_{\mu}~\Theta^{-1}.
\qee
If we use the infinitesimal form of Lorentz transformations
\be
\Lambda_{\mu \nu}= \eta_{\mu \nu} + \epsilon_{\mu \nu}~~~~~~~~\Theta =
1 + \frac{1}{2}\epsilon_{\mu \nu}~I^{\mu \nu}
\qee
it follows that gamma matrices should satisfy the Majorana commutation relation
\cite{majorana}
\be
[\Gamma_{\mu},~I_{\lambda \rho}] = \eta_{\mu \lambda}~\Gamma_{\rho}
- \eta_{\mu \rho}~\Gamma_{\lambda}
\qee
or in components (see formulas (13) in \cite{majorana})
\be
[\Gamma_{0},~a_x] =0~~~~~~[\Gamma_{0},~b_x] =i \Gamma_{x}
\qee

\be
[\Gamma_{x},~a_x] =0~~~[\Gamma_{x},~a_y] =
i\Gamma_z~~~[\Gamma_{x},~a_z] =-i\Gamma_y
\qee

\be
[\Gamma_{x},~b_x] =i\Gamma_{0}~~~~~[\Gamma_{x},~b_y] = 
0~~~~~[\Gamma_{x},~b_z] =0~~~~
\qee
From these equations 
it follows that $\Gamma_{\mu}$ should satisfy the 
equation\footnote{If $\Gamma^{(1)}_{\mu}$ and $\Gamma^{(2)}_{\mu}$ are two
solutions of the equation (\ref{gamma}), then the sum is also a solution and 
if  $\Gamma_{\mu}$ is a solution then using (\ref{gode}) one can see that 
$\Omega \Gamma^{+}_{\mu} \Omega$  is also a solution.}
\be
[[\Gamma_{\mu},~b_{k}],~b_{k}] = 
-\Gamma_{\mu}~~~~~~~~~~~~k=x,y,z. \label{gamma}
\qee
One can also derive that 
\be
\{ \Gamma_{0}~\Gamma_{k}\}=i~[b_{k}~\Gamma^{2}_{0}],
\qee 
thus the anticommutator between $\Gamma_{0}$ and 
$\Gamma_{k}$ essentially depends on the form of the $\Gamma^{2}_{0}$ operator.
These are the most important equations because they allow to find 
gamma matrices when a representation $\Theta$ of the Lorentz algebra
is given. It is an art to choose an appropriate representation $\Theta$ 
in order to have an equation with the necessary properties.
We shall choose $N$ pairs of adjoint representations with $j_{0}= 
1/2,...,N-1/2$ (see Fig.1).

\begin{figure}
\centerline{\hbox{\psfig{figure=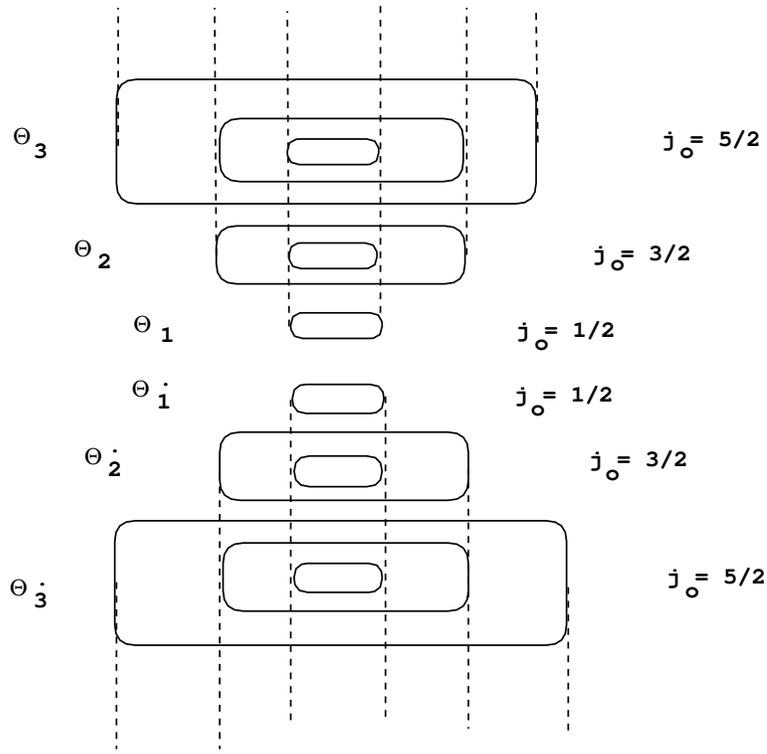,height=10cm,angle=0}}}
\caption[fig1]{The tower of infinite-dimensional representations 
of Lorentz group $\Theta=(\Theta_{\dot{N}},\cdots,
\Theta_{\dot{1}},\Theta_{1},\cdots, \Theta_{N})$. The number of states on the  
spin level j is equal to j+1/2.}
\label{fig1}
\end{figure}

Because $\Gamma_{0}$ commutes with spatial rotations $\vec{a}$
it should have the form
\be
<j,m\vert~\Gamma^{rr'}_{0}~\vert j'm'> = \gamma^{r r'}_{j} \cdot
\delta_{j j'}\cdot \delta_{m m'}~~~~~~~r,r' = \dot{N},...,\dot{1},
1,...,N
\qee
where we consider $N$ pairs of adjoint representations 
$\Theta=(\Theta_{\dot{N}},\cdots,
\Theta_{\dot{1}},\Theta_{1},\cdots, \Theta_{N})$, 
thus $\gamma^{rr'}$ is $2N \times 2N$ matrix which 
should satisfy the equation (\ref{gamma}) for $\mu =0$ and the 
wave function has the form (see Fig.1)
\be
\left( \begin{array}{c}
        \Psi^{r}_{jm}\\
        \Psi^{\dot{r}}_{jm}
\end{array} \right)~~~r=1,...,N~~~~~~j=r-1/2,~r+1/2,...~~~~~~~m=-j,...,j.
\qee 
It should be understood that 
\be
\vec{a}^{~rr'} = \delta^{rr'} \cdot \vec{a}~~~~~~~~\vec{b}^{~rr'} =
\delta^{rr'} \cdot \vec{b}^{~r}.
\qee
One can compute $\Gamma_{k}$ matrices using relation
$\Gamma_{k}=i~(b_{k}~\Gamma_{0}~-~\Gamma_{0}~b_{k})$,~~$k=x,y,z$
$$
<j,m,r~\vert~ \Gamma_{z}~\vert~r+1,j,m> =
i~m~(\lambda^{r}_{j}-\lambda^{r+1}_{j})
\cdot\gamma^{rr+1}_{j}
$$
$$
<j,m,r+1~\vert~ \Gamma_{z}~\vert~r,j,m> =
-i~m~(\lambda^{r}_{j}-\lambda^{r+1}_{j})
\cdot\gamma^{r+1r}_{j}
$$
$$
<j-1,m,r~\vert~ \Gamma_{z}~\vert~ r+1,j,m> =
i~\sqrt{j^2-m^2}\cdot(\varsigma^{r}_{j}\gamma^{rr+1}_{j}-
\gamma^{rr+1}_{j-1}\varsigma^{r+1}_{j})
$$
$$
<j-1,m,r+1~\vert~ \Gamma_{z}~\vert~ r,j,m> =
i~\sqrt{j^2-m^2}\cdot(\varsigma^{r+1}_{j}\gamma^{r+1r}_{j}-
\gamma^{r+1r}_{j-1}\varsigma^{r}_{j})
$$
$$
<j,m,r+1~\vert~ \Gamma_{z}~\vert~ r,j-1,m> =
-i~\sqrt{j^2-m^2}\cdot(
\varsigma^{r}_{j} \gamma^{r+1r}_{j} - \gamma^{r+1r}_{j-1}\varsigma^{r+1}_{j})
$$
\be
<j,m,r~\vert~ \Gamma_{z}~\vert~ r+1,j-1,m> =
-i~\sqrt{j^2-m^2}\cdot(
\varsigma^{r+1}_{j} \gamma^{rr+1}_{j} - \gamma^{rr+1}_{j-1}\varsigma^{r}_{j}).
\qee
It is appropriate to introduce separate notations for diagonal and nondiagonal
parts of $\Gamma_{k}$, we shall define them as  $\tilde {\Gamma_{k}}$ and 
$\tilde {\tilde {\Gamma_{k}}}$. 

\vspace{0.5cm}
We start by analyzing the situation for some lower values of $N$.
A pattern emerges which is then used to construct the solution in 
the general case. 


\section{N pairs of infinite dimensional representations}
\subsection{N=1}
First we shall consider the pair of infinite-dimensional 
adjoint representations 
$\Theta = (~\Theta_{\dot{1}},\Theta_{1}~)$  
\begin{eqnarray}
\Theta_{\dot{1}}~ ---~\Theta_{1}~~~~\\
(1/2;- \lambda)  ~~~~~~~~~  (1/2; \lambda),
\end{eqnarray}
thus the representation is given by matrices
$$
\vert \vert \vec{a}^{rr'}\vert\vert = \left( \begin{array}{c}

        \vec{a},~~0~~\\
        0~~~, \vec{a}
\end{array} \right) ,~~~~\vert \vert \vec{b}^{rr'}\vert\vert 
= \left( \begin{array}{c}
          \vec{b}^{1},~~0~~\\
          0~~~, \vec{b}^{\dot{1}}
\end{array} \right) ,~~~~\Psi = \left( \begin{array}{c}
        \psi_{1}\\
        \psi_{\dot{1}}
\end{array} \right)                 \nonumber
$$ 
with the transition amplitudes 
$$
\lambda^{1}_{j} = - \lambda^{\dot{1}}_{j}
= i ~\frac{1}{2}~\frac{\lambda}
{j(j+1)}~~~~~~~~~j\geq 1/2           \nonumber   
$$
and
$$
\varsigma^{1}_{j} = \varsigma^{\dot{1}}_{j}=
\frac{1}{2} \sqrt{(~1-(\frac{\lambda}
{j})^2~)}~~~~~~~~j \geq 3/2 .               \nonumber 
$$
We are searching for the solution of the equations (\ref{gamma}) in the form of 
Jacoby matrices (\ref{jacob}) (see Appendix A). For $N=1$  
the solution of (\ref{gamma}) is    
\be
\gamma_{j} = \left( \begin{array}{c}
             0~~~ , j + 1/2 \\
             j + 1/2~, ~~~0
\end{array} \right)~~~~~~~~~~~~~~j \geq 1/2
\qee
with the characteristic equation
\be
\gamma^{2}_{j}-(j+1/2)^2 =0
\qee
thus the positive eigenvalues $\epsilon_{j}$ are equal to
\be
\epsilon_{j}=j+1/2~~~~~~~~~~~~~~j\geq 1/2                   \label{spe}
\qee
and grow linearly with j. Therefore the mass spectrum decreases like in 
(\ref{majorana}). The determinant and the trace are equal to
\be
Det~\gamma_{j} = (-1)(j+1/2)^2~~~~~~~~~~Tr~\gamma^{2}_{j}= 
2(j+1/2)^{2}
\qee

\subsection{N=2}

Now we shall take two pairs of adjoint representations
$\Theta = (~\Theta_{\dot{2}},~\Theta_{\dot{1}},
\Theta_{1}~\Theta_{2})$ 
\begin{eqnarray}
\Theta_{\dot{2}}~---~\Theta_{\dot{1}}~ ---~\Theta_{1}~---
\Theta_{2}~~~~~\\
(3/2;- \lambda)~~~~~(1/2;- \lambda)~~~~~~(1/2; 
\lambda)~~~~~(3/2; \lambda),
\end{eqnarray}
then the representation is defined by the matrices of the 
Lorentz algebra

$$
\vert\vert~\vec{a}^{rr'}~\vert\vert = \left( \begin{array}{c}
        \vec{a},~~~0,~~~~0,~~~~0\\
        0,~~~~\vec{a},~~~0,~~~~0\\
        0,~~~~0,~~~~\vec{a},~~~0\\
        0,~~~~0,~~~~0,~~~~ \vec{a}
\end{array} \right),~~\vert\vert~\vec{b}^{rr'}~\vert\vert 
= \left( \begin{array}{c}
        \vec{b}^{2},~~~0,~~~~0,~~~~0\\
        0,~~~~\vec{b}^{1},~~~0,~~~~0\\
        0,~~~~0,~~~~\vec{b}^{\dot{1}},~~~0\\
        0,~~~~0,~~~~0,~~~~ \vec{b}^{\dot{2}}
\end{array} \right),~~\Psi = \left( \begin{array}{c}
        \psi_{2}\\
        \psi_{1}\\
        \psi_{\dot{1}}\\
        \psi_{\dot{2}}
\end{array} \right)                     \nonumber
$$
where the  transition amplitudes are
$$
\lambda^{1}_{j} = - \lambda^{\dot{1}}_{j}
= i ~\frac{1}{2}~\frac{\lambda}
{j(j+1)}~~~~~~~~~j\geq 1/2                           \nonumber
$$
$$
\lambda^{2}_{j} = - \lambda^{\dot{2}}_{j}
= i ~\frac{3}{2}~\frac{\lambda}
{j(j+1)}~~~~~~~~~j\geq 3/2                            \nonumber
$$                                         
and

$$
\varsigma^{1}_{j} = \varsigma^{\dot{1}}_{j}=
\frac{1}{2} \sqrt{(~1-(\frac{\lambda}
{j})^2~)}~~~~~~~~j \geq 3/2                               \nonumber
$$

$$
\varsigma^{2}_{j} = \varsigma^{\dot{2}}_{j}=
\frac{1}{2} \sqrt{(~1- \frac{2}
{j^{2}-1/4})} ~~\sqrt{(~1-(\frac{\lambda}
{j})^2~)}~~~~~~~~j \geq 5/2 .                              \nonumber
$$
Again we are searching for a solution of equations (\ref{gamma}) in 
the form of the second matrix (\ref{jacob}) presented in Appendix A. For $N=2$ 
the representation  of the solution of (\ref{gamma}) is  

\be
\gamma_{1/2}= \left( \begin{array}{c}
  0,~~~1\\
  1,~~~0
\end{array} \right),~~~~\gamma_{j} = \left( \begin{array}{c}
0,~~~~~~~i\sqrt{\frac{3}{4}((j^2 +j)/3-1/4)},~~~~~~~~~~~0,~~~~~~~~~~~0\\
i\sqrt{\frac{3}{4}((j^2 +j)/3-1/4)},~~~~~~0,~~~~~~~~~~~j+1/2,~~~~~~~~0\\
0,~~~~~~~~~j+1/2,~~~~~~~~~~0,~~~~~~i\sqrt{\frac{3}{4}((j^2 +j)/3-1/4)}\\
0,~~~~~~~~~~~0,~~~~~~~~~~~i\sqrt{\frac{3}{4}((j^2 +j)/3-1/4)},~~~~~~~0
\end{array} \right)
\qee
where $j \geq 3/2$ and the corresponding determinants and traces are equal to
\be
Det~\gamma_{1/2}=-1,~~~~Tr~\gamma^{2}_{1/2}=2
\qee
\be
Det~\gamma_{j} =\frac{1}{16}(j^2 +j -3/4)^2 ~~~Tr~\gamma^{2}_{j}= 
j^2 + j + 5/4~~~~~j\geq 3/2
\qee
Characteristic equations for these matrices are:
\begin{eqnarray}
\gamma^{2}_{1/2} -1=0,~~~~~~~~~~~~~\\
\gamma^{4}_{j} -\frac{1}{2}(j^2+j+\frac{5}{4})
\gamma^{2}_{j} + \frac{1}{16}(j^{2}+j-\frac{3}{4})^2 =0~~~~~~j\geq 3/2
\end{eqnarray}
and the eigenvalues satisfy the relation
\be
Det~\gamma_{j}=\epsilon_{1}^2\epsilon_{2}^2~~~Tr~\gamma^{2}_{j}=
2(\epsilon_{1}^2 + \epsilon_{2}^2)~~j\geq 3/2.
\qee 
The positive eigenvalues $\epsilon_{j}$ in the given case  are:
\begin{eqnarray}
1~~~~~~~~~~~~~~~~~~~~~~~~ j=1/2\\
\frac{1}{2}(j-1/2)~~~~~\frac{1}{2}(j+3/2)~~~~~~~j\geq 3/2    \label{spec}
\end{eqnarray}
and they grow linearly with $j$ as it was in the previous case $N=1$,
but what is more important, the coefficient of proportionality drops by half
in (\ref{spec}), compared with (\ref{spe}) and  we get an eigenvalue 
$\epsilon_{3/2}$ which is less than unity.

\subsection{N=3}
The solution in this case $\Theta = 
(~\Theta_{\dot{3}}~\Theta_{\dot{2}},~\Theta_{\dot{1}},
\Theta_{1}~\Theta_{2}~\Theta_{3})$ has the form
\be
\gamma_{1/2}= \left( \begin{array}{c} ~~~1\\1~~~ \end{array} 
\right),~~~~\gamma_{3/2} = \left( \begin{array}{c}
0,~~~~~~~i\sqrt{\frac{8}{9}},~~~~~~~~~~~~~~~~~~\\
i\sqrt{\frac{8}{9}},~~~~~~0,~~~~~~~~~2~~~~~~~~~\\~~~~~~~~~~~2
,~~~~~~~~0,~~~~~~i\sqrt{\frac{8}{9}}\\~~~~~~~~~~~~~~~~~~~~i
\sqrt{\frac{8}{9}},~~~~~~~0
\end{array} \right)
\qee

\be
\gamma_{j} = \left(\begin{array}{l}
~~~~0,~~~i\sqrt{\frac{5}{9}((j^2 +j)/15-1/4)},~~~~~~~~~~~~~~~~~~~~~~~\\
i\sqrt{\frac{5}{9}((j^2 +j)/3-1/4)},~~~~~~0,~~~~i\sqrt{\frac{8}{9}
((j^2 +j)/3-1/4)},~~~~~~~~~~\\~~~~~~~~~~i\sqrt{\frac{8}{9}((j^2 +j)/3-1/4)}
,~~~~~~0,~~~~~~~~j+1/2
,~~~~~~~~~~~~~~~~~~~~\\~~~~~~~~~~~~~~j+1/2,~~~~~~~~~~0
,~~~~~~i\sqrt{\frac{8}{9}((j^2 +j)/3-1/4)}\\~~~~~~~~~~~~~~~~~~~~~i
\sqrt{\frac{8}{9}((j^2 +j)/3-1/4)},~~~~~~~~0
,~~~~~~~~~i\sqrt{\frac{5}{9}
((j^2 +j)/15-1/4)}\\~~~~~~~~~~~~~~~~~~~~~~~~~~~~~i
\sqrt{\frac{5}{9}((j^2 +j)/15-1/4)},~~~~~~~0
\end{array} \right)
\qee
where $j \geq 5/2$ and the corresponding determinants and traces are equal to
$$
Det~\gamma_{1/2}=-1,~~~~Tr~\gamma^{2}_{1/2}=2           \nonumber
$$
$$
Det~\gamma_{3/2} = (1-1/9)^2~~~~Tr~\gamma^{2}_{3/2}= 40/9         \nonumber
$$
\be
Det~\gamma_{j} = (-1/27^{2})(j +1/2)^{2}
(j^2 + j - 15/4)^2~~~Tr~\gamma^{2}_{j}=
\frac{2}{3}(j^2 + j + 35/12)~~~~j\geq 5/2                    
\qee
In this case the characteristic equations are:
\begin{eqnarray}
\gamma^{2}_{1/2}-1=0~~~~~~~~~~~~~\nonumber\\
((\gamma_{3/2}-1)^{2} -1/9)~((\gamma_{3/2}+1)^{2} -1/9) 
=0~~~~~~~~~~~~~\nonumber\\
(~\gamma^{3}_{j} +(j+\frac{1}{2})\gamma^{2}_{j} + 
\frac{1}{3}(j^{2}+j-\frac{13}{12})\gamma_{j}+
\frac{1}{27}(j+\frac{1}{2})(j^{2}+j-\frac{15}{4}
)~)~~~~~~~~~~~~~~~~~~\nonumber\\
(~\gamma^{3}_{j} -(j+\frac{1}{2})\gamma^{2}_{j} +
\frac{1}{3}(j^{2}+j-\frac{13}{12})\gamma_{j}-
\frac{1}{27}(j+\frac{1}{2})(j^{2}+j-\frac{15}{4})~) =0~~j\geq 5/2
\end{eqnarray}
and eigenvalues satisfy the relations
\be
Det~\gamma_{j} = -\epsilon_{1}^2\epsilon_{2}^2
\epsilon_{3}^2~~~~~Tr~\gamma^{2}_{j}=
2(\epsilon_{1}^2 + \epsilon_{2}^2+ \epsilon_{3}^2)~~~~j\geq 5/2.
\qee 
The positive eigenvalues are equal to (see Fig.2)
\begin{eqnarray}
1~~~~~~~~~~~~~~~~~~~~~~~~~~~~~~~~~~~~~~~~~~~j=1/2\nonumber\\
1-1/3~~~~~~~~~~~~~~~~~1+1/3~~~~~~~~~~~~~~~~~~~~~~~~j=3/2\nonumber\\
\frac{1}{3}(j-3/2)~~~~~~~~~\frac{1}{3}(j+1/2)~~~~~~~~~~~~\frac{1}{3}
(j+5/2)~~~~~~~~j \geq 5/2.                                    \label{sp}
\end{eqnarray}
This solution demonstrates that when we consider three pairs of 
infinite-dimensional representations
the rate of growth of the positive 
eigenvalues as a function of $j$ drops three times in (\ref{sp}), compared with
(\ref{spe}) and (\ref{spec}). {\it This actually means 
that by increasing the number of representations we can 
slow down the growth of the eigenvalues}. To see how it happens let us consider 
$N$ pairs of adjoint representations.

\begin{figure}
\centerline{\hbox{\psfig{figure=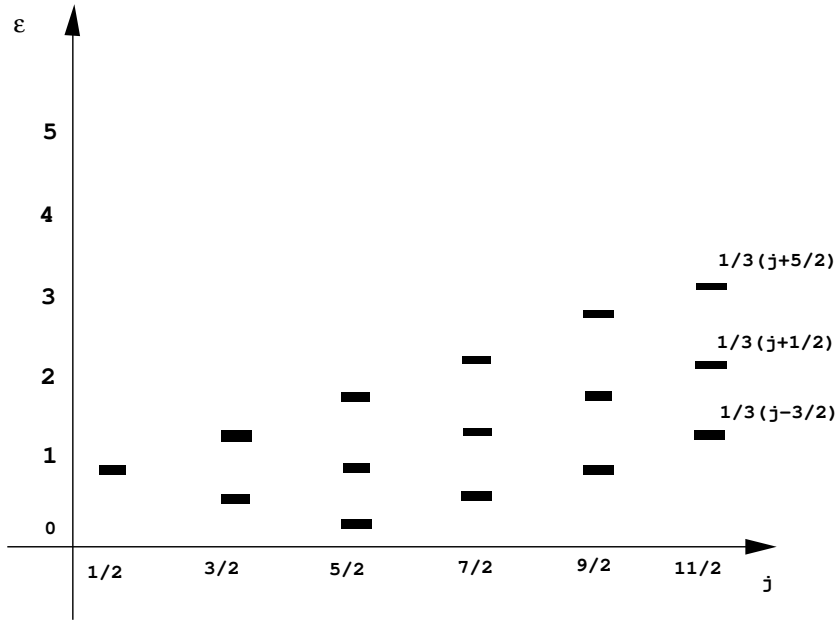,height=10cm,angle=0}}}
\caption[fig2a]{The positive part of spectrum of the $\Gamma_{0}$ matrix for N=3. 
One can clearly see the appearance of eigenvalues which are less than unity, 
$\epsilon < 1$.}
\label{fig2a}
\end{figure}
\begin{figure}
\centerline{\hbox{\psfig{figure=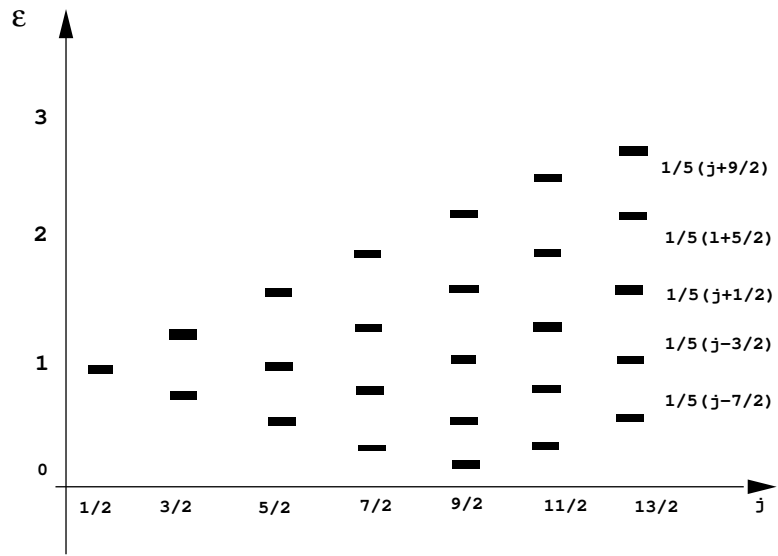,height=10cm,angle=0}}}
\caption[fig2b]{The positive part of spectrum of the $\Gamma_{0}$ matrix for N=5.
The number of states with $\epsilon < 1$ increases. As N tends to infinity the 
spectral cone $j= [1/2,N-1/2]$ becomes more "narrow" and the eigenvalues 
are concentrated around $\epsilon=1$.}
\label{fig2b}
\end{figure}

\subsection{N pairs of adjoint representations}

In this case we shall take $\Theta=(\Theta_{\dot{N}},\cdots, 
\Theta_{\dot{1}},\Theta_{1},\cdots, \Theta_{N})$ which can be rewritten 
also in the form (\ref{repres})
\be
(N-1/2;- \lambda),~~~\cdots~~~,(1/2
;- \lambda)~,~(1/2; \lambda),~~~\cdots~~~,(N-1/2; \lambda), \label{basicrep}
\qee
the wave function has the form
$$
\Psi = \left( \begin{array}{c}
         \Psi^{r}_{jm}\\
         \Psi^{\dot{r}}_{jm}        
\end{array} \right)~~~~~~~~r=
1,...,N;~~~~~~j=r-1/2,~r+1/2,...;~~~~~~~m=-j,...,j     \nonumber
$$
Then the transition amplitudes are equal to 
$$
\lambda^{r}_{j} = - \lambda^{\dot{r}}_{j}
= i ~\frac{r - 1/2}{j(j+1)}~\lambda~~~~~~~~~~~~~~~j\geq r-1/2 \nonumber
$$
and
$$
\varsigma^{r}_{j} = \varsigma^{\dot{r}}_{j}=
\frac{1}{2} \sqrt{(1-\frac{r^2-r}{j^2-1/4})(1-(\frac{\lambda}
{j})^2)}~~~~~~j \geq r+1/2          \nonumber
$$
where $~~~r=1,...,N$. 

\vspace{.5cm}
Our basic solution ( {\it B-solution}) of the equation (\ref{gamma}) for the 
$\Gamma_{0}$ in 
this general case is a $~2N\times 2N~$ Jacoby matrix with the following 
nonzero elements
\be
\gamma^{r+1~r}_{j}=\gamma^{r~r+1}_{j} =
\gamma^{\dot{r+1}~\dot{r}}_{j}= \gamma^{\dot{r}~\dot{r+1}}_{j} = 
i~ \sqrt{(1-\frac{r^2}{N^2})~(\frac{j^2 + j}
{4r^2-1} -\frac{1}{4})}~~~~~j\geq r+1/2                   \label{solution} 
\qee
\be
\gamma^{1~\dot{1}}_{j} = \gamma^{\dot{1}~1}_{j} = j+1/2
\qee
here $r=1,...,N-1$. 
The gamma matrices for the first few values of $j$ are equal to 
\vspace{.5cm}
$$
\gamma_{1/2} = \left( \begin{array}{c}
  0,~~~~~~1\\
  1,~~~~~~0
\end{array} \right),~~ \gamma_{3/2} = \left( \begin{array}{c}
  0,~~~~~~~~~~~~~i\sqrt{1-1/N^2},~~~~~~~~~~~~~~~~~~~~~~~\\
  i\sqrt{1-1/N^2},~~~~~0,~~~~~~2
,~~~~~~~~~~~~~~~~~~~~~~~\\~~~~~~~~~~~~~~~~~~~~~~~~2
,~~~~~~0,~~~~~~~~~~i
\sqrt{1-1/N^2}\\~~~~~~~~~~~~~~~~~~~~~~~~~~~~i
\sqrt{1-1/N^2},~~~~~~~~~~~0
\end{array} \right),         \nonumber
$$
$$
\gamma_{5/2} = \left( \begin{array}{c}
  0,~~~~~~i\sqrt{(1-4/N^2)(1/3)},~~~~~~~~~~~~~~~~~~~~~~\\
  i\sqrt{(1-4/N^2)(1/3)},~~~~~0,~~~~~~~~~~i
\sqrt{(1-1/N^2)(8/3)},~~~~~~~~~~~~~~~~~\\~~~~~~~~~~~~~i
\sqrt{(1-1/N^2)(8/3)},~~~~0,~~~~~~~~~~3
,~~~~~~~\\~~~~~~~~~~~~~~~~~~~~~~~~~~~~~~~~~~~~~~~~~3
,~~~~~~~~~~0,~~~~~i
\sqrt{(1-1/N^2)(8/3)}
,~~~~~~~~~\\~~~~~~~~~~~~~~~~~~~~~~~~~~i
\sqrt{(1-1/N^2)(8/3)},~~~0
,~~~i\sqrt{(1-4/N^2)(1/3)}
\\~~~~~~~~~~~~~~~~~~~~~~~~~~~~~~~~~~~~~~~i
\sqrt{(1-4/N^2)(1/3)},~~~0
\end{array} \right),            \nonumber
$$

\vspace{.5cm}
\noindent
and they grow in size with $j$ until $j=N - 1/2$, for greater  $j$ 
the size of the matrix $\gamma_{j}$ remains the same and is equal to
$2N \times 2N$. 
The determinant of the $\gamma_{j}$ matrix
for $j$ in the interval $[3/2,~7/2,...,N - 1/2]$ is equal to 
\be
Det~\gamma_{j} = 
(1-1/N^{2})^{2}~\cdots~(1-(j-1/2)^{2}/N^{2})^{2}
\qee
and for $j$ in the interval $[1/2,~5/2,..., N - 1/2]$ is 
\be
Det~\gamma_{j} = (-1)
(1-4/N^{2})^{2}~\cdots~(1-(j-1/2)^{2}/N^{2})^{2}.
\qee
From the determinant and the trace of the gamma matrix it follows that 
\be
\epsilon_{1}^2 \cdot ... \cdot \epsilon_{j+1/2}^{2} = \vert Det~\gamma_{j}
\vert~~~~~~~~\epsilon_{1}^{2}
 + ...+\epsilon_{j+1/2}^{2} = \frac{1}{2}Tr~\gamma^{2}_{j}~~~~j\geq (N - 1/2).
\qee
The characteristic equations for these matrices are:
\begin{eqnarray}
\gamma^{2}_{1/2}-1=0,\nonumber\\
((\gamma_{3/2}-1)^2 - 1/N^2)~((
\gamma_{3/2}+1)^2 - 1/N^2) =0,\nonumber\\
(\gamma^{2}_{5/2}-1)~((\gamma_{5/2}-1)^2 - 4/N^2)~((
\gamma_{5/2}+1)^2 - 4/N^2) =0,\nonumber\\
.~.~.~.~.~.~.~.~.~.~.~.~.~.~.~.~.~.~.~.~.~.~.~.~.~.~.~.~.~.~.~.~.~.~.~.
~.~.~.~.~.~.\nonumber\\
\cdots (\gamma^{2}_{j}-(1 + (j-5/2)/N)^2) ~(
\gamma^{2}_{j}-(1 + (j-1/2)/N)^2) =0,
\end{eqnarray}
where in the last equation $j \geq N-1/2$.
The positive eigenvalues $\epsilon_{j}$ can be now found  
$$
        1~~~~~~~~~~~~~~~~~~~~~~~~~~~~~j=1/2  \nonumber
$$
$$
1-1/N~~~~~~~~1+1/N~~~~~~~~~~~~~~~~~~~~~~~~~~~~~~~~j=3/2      \nonumber
$$
$$
1-2/N~~~~1~~~1+2/N~~~~~~~~~~~~~~~~~~~~~~~~~~~~~~~~j=5/2       \nonumber
$$
$$
1-3/N~~~~1-1/N~~~~~1+1/N~~~~1+3/N~~~~~~~~~~~~~~~~~~~~~~~~~j=7/2   \nonumber
$$
$$
1-4/N~~~~1-2/N~~1~~1+2/N~~~~1+4/N~~~~~~~~~~~~~~~~~~~~~~~~~j=9/2    \nonumber
$$
~.~.~.~.~.~.~.~.~.~.~.~.~.~.~.~.~.~.~.~.~.~.~.~.~.~.~.~.~.~.~.
~.~.~.~.~.
\vspace{.5cm}
\be
\cdots~~,1 + (j-5/2)/N,~~~1+(j-1/2)/N~~~~~~~~~j\geq N-1/2
\qee
On Fig.2,3 one can see the spectrum of the matrix $\gamma_{j}$ as a 
functin of $j$.

The number of states with angular momentum $j$  grows as
$j+1/2$ and  this takes place up to spin $j=N-1/2$. For the higher spins 
$j \geq N-1/2$ the number of states remains constant and is equal to $N$
(see Fig.2,3). Thus we see that the coefficient of 
proportionality drops N times and many eigenvalues are less than unity. 
The mass spectrum is bounded from bellow only if all eigenvalues are less than 
unity.

\section{Nonhermitian Solution of $\Gamma_{0}$. $N \rightarrow \infty$.}

In the limit $N \rightarrow \infty$ our solution (\ref{solution})
is being reduced to the form
\be
\gamma^{r+1~r}_{j}=\gamma^{r~r+1}_{j} =
\gamma^{\dot{r+1}~\dot{r}}_{j}= \gamma^{\dot{r}~\dot{r+1}}_{j} = 
i~ \sqrt{(\frac{j^2 + j}
{4r^2-1} -\frac{1}{4})}~~~~~j\geq r+1/2  
\qee
\be
\gamma^{1~\dot{1}}_{j} = \gamma^{\dot{1}~1}_{j} = j+1/2
\qee
where $r=1,2,...$. As it is easy to see from the previous formulas, 
all eigenvalues $\epsilon_{j}$ 
tend to unity when number of representations
$N \rightarrow \infty$.
The characteristic equation which is satisfied by the gamma matrix in 
this limit is
\be
(\gamma_{j}^2 -1)^{j+1/2} = 0 ~~~~~~~~~~j=1/2,~3/2,~,5/2,\cdots
\qee
with all eigenvalues $\epsilon_{j} = \pm 1$ (see Fig.4). 

\begin{figure}
\centerline{\hbox{\psfig{figure=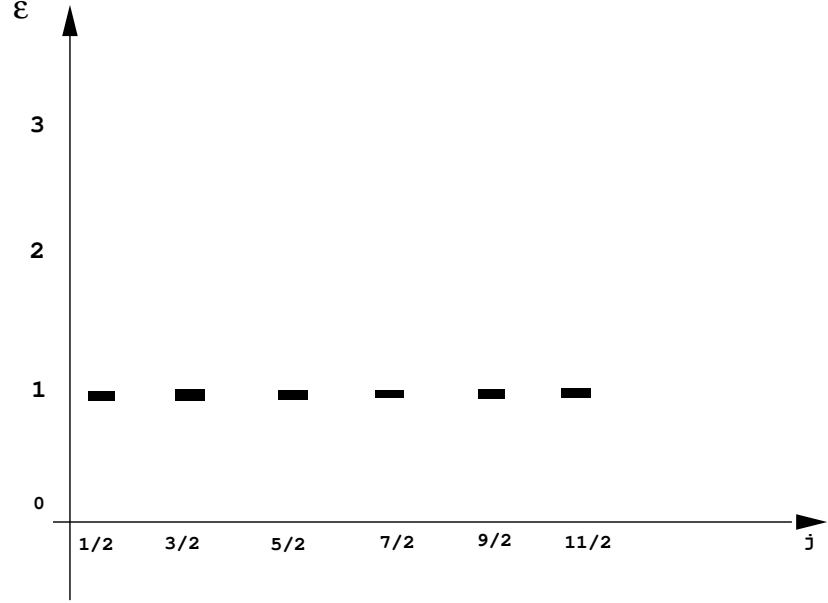,height=10cm,angle=0}}}
\caption[fig2c]{The positive part of the spectrum of the $\Gamma_{0}$ matrix 
for $N=\infty$.}
\label{fig2c}
\end{figure}

The determinant
and the trace are equal to
\be
Det~\gamma_{j} = \pm 1~~~~~Tr~\gamma^{2}_{j} = 2j +1,
\qee
thus
\be
\epsilon_{1}^2 \cdot ...\cdot\epsilon_{j+1/2}^2= 1~~~~~~\epsilon_{1}^2 + 
...+\epsilon_{j+1/2}^2 = j+1/2
\qee
The matrix $\Omega~\Gamma_{0}$ has the characteristic equation
\be
(\omega_{j}~\gamma_{j} - 1)^{2j + 1} = 0
\qee
with all eigenvalues equal to $\rho_{j} = +1$, therefore introducing 
\be
\omega_{j}~\gamma_{j}-1 = \sigma_{j}
\qee
we have an important relation
\be
\sigma^{2j+1}_{j} = 0
\qee
and 
\be
\gamma_{j} = \omega_{j}(~1 + ~ \sigma_{j}~)
\qee
The $\sigma_{j}$ defines the algebra of elements
\be
\sigma_{j},~\sigma^{+}_{j},~\sigma^{2}_{j},~\sigma^{+~2}_{j},...,
\sigma^{2j}_{j},~\sigma^{+~2j}_{j}
\qee
with the property
\be
\sigma^{2j+1}_{j} = \sigma^{+~2j+1}_{j} =0
\qee

Thus the matrix $\Omega~\Gamma_{0}$ is positive definite and all its 
eigenvalues are equal to one, but the important relations
\be
\Omega~\Gamma_{0} \neq \Gamma^{+}_{0}~\Omega~~~~~~\Gamma^{+}_{0} \neq
\Gamma_{0}
\qee
are not held and therefore the 
Hamiltonian is not Hermitian. 
In the next section we shall find the Hermitian solution for $\Gamma_{0}$
using the fact 
that one can change the phases of the matrix elements without disturbing 
its determinant.

\section{Hermitian solution for $\Gamma_{0}$}
The Hermitian solution ({\it H-solution}) of (\ref{gamma}) for $\Gamma_{0}$
can be found as a phase modification of our basic {\it B-solution} (\ref{solution})
\be
\gamma^{r+1~r}_{j}=-\gamma^{r~r+1}_{j} =
-\gamma^{\dot{r+1}~\dot{r}}_{j}= \gamma^{\dot{r}~\dot{r+1}}_{j} =
i~ \sqrt{(\frac{j^2 + j}{4r^2-1} -\frac{1}{4})}~~~~~j\geq r+1/2   \label{herso}
\qee
\be
\gamma^{1~\dot{1}}_{j} = \gamma^{\dot{1}~1}_{j} = j+1/2
\qee
which for low values of $j$ is:
$$
\gamma_{1/2} = \left( \begin{array}{c}
  0,~~~1\\
  1,~~~0
\end{array} \right),~~ \gamma_{3/2} = \left( \begin{array}{c}
  0,~~~i,~~~~~~~~~\\
  -i,~~0,~~~2,~~~~\\~~~~~2,~~~0,~~~i\\~~~~~~~~~-i,~~~0
\end{array} \right),                      \nonumber
$$
$$
\gamma_{5/2} = \left( \begin{array}{c}~~~~~0,~~~~~i
\sqrt{1/3},~~~~~~~~~~~~~~~~~~~~~~~~~~~~~~~~~~~~~~\\
-i\sqrt{1/3},~~~0,~~~i\sqrt{8/3}
,~~~~~~~~~~~~~~~~~~~~~~~~~~~~~\\~~~~~~~~~~~-i\sqrt{8/3}
,~~~0,~~~~3
,~~~~~~~~~~~~~~~~~~~~~~~~~~~\\~~~~~~~~~~~~~~~~~~~~~~~~~~~3
,~~~~0,~~~~i\sqrt{8/3}
,~~~~~~~~~~~\\~~~~~~~~~~~~~~~~~~~~~~~~~~~~-i
\sqrt{8/3},~~~0,~~~~~i
\sqrt{1/3}\\~~~~~~~~~~~~~~~~~~~~~~~~~~~~~~~~~~~~~~~-i
\sqrt{1/3},~~~~0~~~~~\end{array} \right),...           \nonumber
$$
These matrices are Hermitian $\Gamma^{+}_{0}=\Gamma_{0}$, but
the characteristic equations are more complicated now:
\begin{eqnarray}
\gamma^{2}_{1/2}-1=0,\nonumber\\
(\gamma^{2}_{3/2}- 2\gamma_{3/2}-1)~(
\gamma^{2}_{3/2}+ 2\gamma_{3/2}-1) =0,\nonumber\\
(\gamma^{2}_{5/2}-1)~(\gamma^{2}_{5/2} - 4\gamma_{5/2}+1)~(
 \gamma^{2}_{5/2} + 4\gamma_{5/2}+1)=0,\nonumber\\
(\gamma^{4}_{7/2}+4\gamma^{3}_{7/2}-
6\gamma^{2}_{7/2}-4\gamma_{7/2}+1 )
(\gamma^{4}_{7/2}-4\gamma^{3}_{7/2}-
6\gamma^{2}_{7/2}+4\gamma_{7/2}+1 )=0,\nonumber\\
(\gamma^{2}_{9/2}-1)~(\gamma^{4}_{9/2}+4\gamma^{3}_{9/2}-
14\gamma^{2}_{9/2}+4\gamma_{9/2}+1 )~(\gamma^{4}_{9/2}-4\gamma^{3}_{9/2}-
14\gamma^{2}_{9/2}-4\gamma_{9/2}+1 )=0,\nonumber\\
.~.~.~.~.~.~.~.~.~.~.~.~.~.~.~.~.~.~.~.~.~.~.~.~.~.~.~.~.~.~.~.~.~.~.~.~.~.~.~.~.~.~.~.~.~
.~.~.~.~.~.~.~.~.~.~.~.~.~.\nonumber\\
\end{eqnarray}
These  polynomials $p(\epsilon)$ have the reflective symmetry and are even
\be
p_{j}(\epsilon)~= ~\epsilon^{2j+1}~p_{j}(1/\epsilon)
\qee
\be
p_{j}(-\epsilon)~= ~p_{j}(\epsilon)
\qee
therefore 
if $\epsilon_{j}$ is a solution then $1/\epsilon_{j}$,$-\epsilon_{j}$
and $-1/\epsilon_{j}$ are also solutions.
Computing the traces and determinants of these matrices one can get 
the following general relation for the eigenvalues
\be
\epsilon_{1}^2 \cdot ...\cdot\epsilon_{j+1/2}^2= 1~~~~~~~~\epsilon_{1}^2 + 
...+\epsilon_{j+1/2}^2 = j(2j+1).
\qee
The eigenvalues $\epsilon_{j}$ can be now found
$$
~~~~~~~~~~~~~~~~~~~1~~~~~~~~~~~~~~~~~~~~~~~~~~~~~~~~~~~j=1/2       \nonumber    
$$
$$
~~~~~~~~~~~~~~~~~\sqrt{2}-1~~~~~~~~\sqrt{2}+1~~~~~~~~~~~~~~~~~~~~~~~~~~~~~~~~~~j=3/2
\nonumber
$$
$$
~~~~~~~~~~~~~~~~~2-\sqrt{3}~~~~1~~~~2+\sqrt{3}~~~~~~~~~~~~~~~~~~~~~~~~~~~~~~~~~~j=5/2
\nonumber
$$
$$
\sqrt{(\sqrt{2}+1)^2 +1} - (\sqrt{2}+1)~~~~~~~~\sqrt{(\sqrt{2}-1)^2 +1}
- (\sqrt{2}-1)~~~~~~~~~~~~~j=7/2\nonumber
$$
$$
\sqrt{(\sqrt{2}-1)^2 +1} + (\sqrt{2}-1)~~~~~~~~\sqrt{(\sqrt{2}+1)^2 +1} + 
(\sqrt{2}+1)~~~~~~~~~~~~~j=7/2\nonumber
$$
$$
\sqrt{5}+1 - \sqrt{(\sqrt{5}+1)^2 -1}~~~~~~~~\sqrt{5}-1 - 
\sqrt{(\sqrt{5}-1)^2 -1}~~~~~~~~~~~~~j=9/2\nonumber 
$$
$$
~~~~~~~~~~~~~~~~~~~~~~~~~~~~~~~~~1~~~~~~~~~~~~~~~~~~~~~~~~~~~~~~~~~~~~~~~~~~~~~~~~~j=9/2\nonumber
$$
$$
\sqrt{5}-1 + \sqrt{(\sqrt{5}-1)^2 -1}~~~~~~~~\sqrt{5}+1 + 
\sqrt{(\sqrt{5}+1)^2 -1}~~~~~~~~~~~~~j=9/2            \nonumber    
$$
\be
.~.~.~.~.~.~.~.~.~.~.~.~.~.~.~.~.~.~.~.~.~.~.~.~.~.~.~.~.~.~.~.~.~.~.~.\label{osp}
\qee
The changes of the phases in the matrix elements (\ref{herso}) 
result into different behaviour of eigenvalues (see Fig.5).

\begin{figure}
\centerline{\hbox{\psfig{figure=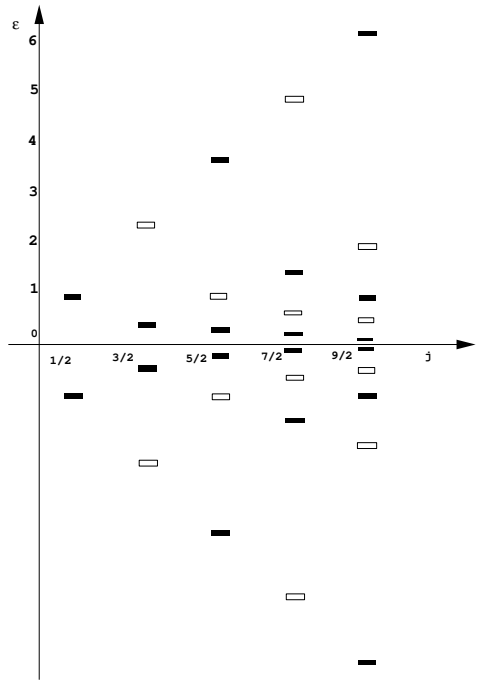,height=10cm,angle=0}}}
\caption[fig3]{The spectrum of the $\Gamma_{0}$ matrix for the 
Hermitian solution (H-solution).  
The quasilinear trajectories have different slope.
One can observe also approximate hyperbolic curves on which lie the 
eigenvalues from different trajectories. The symmetric solution 
($\Sigma$-solution) has the same spectrum.}
\label{fig3}
\end{figure}

\vspace{.5cm}
The matrix $\Omega~\Gamma_{0}$ has again the 
characteristic equation
\be
(\omega_{j}~\gamma_{j} - 1)^{2j + 1} = 0
\qee
and all eigenvalues are equal to one. Thus again the  
matrix $\Omega~\Gamma_{0}$ is positive definite 
because all eigenvalues are equal to one, but the important relation
\be
\Omega~\Gamma_{0} \neq \Gamma^{+}_{0}~\Omega
\qee
does not hold. Introducing 
\be
\omega_{j}~\gamma_{j}-1 = ~ \tau_{j}
\qee
we have the relation
\be
\tau_{j}^{2j+1} = 0
\qee
and 
\be
\gamma_{j} = \omega_{j}(~1 + ~\tau_{j})
\qee
The $\tau_{j}$ defines the algebra of elements
\be
\tau_{j},~\tau^{+}_{j},~\tau^{2}_{j},~\tau^{+~2}_{j},...,
\tau^{2j}_{j},~\tau^{+~2j}_{j}
\qee
with the property
\be
\tau^{2j+1}_{j} = \tau^{+~2j+1}_{j} =0.
\qee

\section{Real and symmetric solution for $\Gamma_{0}$}

The  solution of (\ref{gamma}) for $\Gamma_{0}$ with necessary properties
can be found by using our basic solutions (\ref{solution}) rewritten with
arbitrary phases of the matrix elements and then by requiring that $\Gamma_{0}$
should be Hermitian $\Gamma^{+}_{0} = \Gamma_{0}$ and should satisfy the 
relations $\Gamma^{+}_{0}~\Omega = \Omega~\Gamma_{0}$. The symmetric 
$\Sigma$-solution is
\be
\gamma^{r+1~r}_{j}=~\gamma^{r~r+1}_{j} =
\gamma^{\dot{r+1}~\dot{r}}_{j}= \gamma^{\dot{r}~\dot{r+1}}_{j} =
 \sqrt{(\frac{j^2 + j}{4r^2-1} -\frac{1}{4})}~~~~~j\geq r+1/2    \label{realsym}
\qee
\be
\gamma^{1~\dot{1}}_{j} = \gamma^{\dot{1}~1}_{j} = j+1/2
\qee
which for the low values of $j$ is:
\be
\gamma_{1/2} = \left( \begin{array}{c}
  0,~~~1\\
  1,~~~0
\end{array} \right),~~ \gamma_{3/2} = \left( \begin{array}{c}
  0,~~~1,~~~~~~~~~\\
  1,~~0,~~~2,~~~~\\~~~~~2,~~~0,~~~1\\~~~~~~~~~1,~~~0
\end{array} \right), \label{real}
\qee
\be
\gamma_{5/2} = \left( \begin{array}{c}~~~~~0
,~~~~~\sqrt{1/3},~~~~~~~~~~~~~~~~~~~~~~~~~~~~~~~~~~~~~~\\
\sqrt{1/3},~~~0,~~~\sqrt{8/3}
,~~~~~~~~~~~~~~~~~~~~~~~~~~~~~\\~~~~~~~~~~~\sqrt{8/3}
,~~~0,~~~~3
,~~~~~~~~~~~~~~~~~~~~~~~~~~~\\~~~~~~~~~~~~~~~~~~~~~~~~~~~3
,~~~~0,~~~~\sqrt{8/3}
,~~~~~~~~~~~\\~~~~~~~~~~~~~~~~~~~~~~~~~~~~\sqrt{8/3}
,~~~0,~~~~~\sqrt{1/3}
\\~~~~~~~~~~~~~~~~~~~~~~~~~~~~~~~~~~~~~~~\sqrt{1/3}
,~~~~0~~~~~\end{array} \right),...  \nonumber
\qee
In this case we have Hermitian matrix $\Gamma^{+}_{0} = \Gamma_{0}$ 
which has the desired property 
\be
\Gamma^{+}_{0}~\Omega = \Omega~\Gamma_{0}.
\qee
This  means that the charge density is equal to $\rho =\Omega~\Gamma_{0}$.
In addition all gamma matrices now have this property (\ref{holdc})
\be
\Gamma^{+}_{k}~\Omega = \Omega~\Gamma_{k}~~~~~ k=x,y,z
\qee
which follows from the definition of $\Gamma_{k}= i [b_{k}~\Gamma_{0}]$ and
equation (\ref{gode}) $\Omega~b_{k}~ = b^{+}_{k}~\Omega$.

The characteristic equations for the first few values of $j$  are
\begin{eqnarray}
\gamma^{2}_{1/2}-1=0,\nonumber\\
(\gamma^{2}_{3/2}-2\gamma_{3/2} - 1)~(\gamma^{2}_{3/2}+2\gamma_{3/2} - 
1)=0,\nonumber\\
(\gamma^{2}_{5/2}-1)~(\gamma^{2}_{5/2} - 4\gamma_{5/2} +1)~(
\gamma^{2}_{5/2} + 4\gamma_{5/2} +1)=0,\nonumber\\
(\gamma^{2}_{7/2}+2(\sqrt{2}-1)\gamma_{7/2}-1)~(
\gamma^{2}_{7/2}-2(\sqrt{2}+1)\gamma_{7/2}-1)~~~~~~~\nonumber\\
(\gamma^{2}_{7/2}+2(\sqrt{2}+1)\gamma_{7/2}-1)~(
\gamma^{2}_{7/2}-2(\sqrt{2}-1)\gamma_{7/2}-1)=0,\nonumber\\
(\gamma^{2}_{9/2}-1)~(\gamma^{2}_{9/2}+2(\sqrt{5}-1)\gamma_{9/2}+1)~(
\gamma^{2}_{9/2}-2(\sqrt{5}+1)\gamma_{9/2}+1)~~~~~~~\nonumber\\
(\gamma^{2}_{9/2}+2(\sqrt{5}+1)\gamma_{9/2}+1)~(
\gamma^{2}_{9/2}-2(\sqrt{5}-1)\gamma_{9/2}+1)=0,\nonumber\\
.~.~.~.~.~.~.~.~.~.~.~.~.~.~.~.~.~.~.~.~.~.~.~.~.~.~.~.~.~.~.~.~.~.~.~.~.~.~.~.~.~.~.~.~
\end{eqnarray}
and the spectrum (\ref{osp}) (see Fig.5) is the same for the Hermitian 
H-solution and symmetric $\Sigma$-solution.
The corresponding characteristic equations for the matrices $\rho_{j}$ are
\begin{eqnarray}
(\rho_{1/2} - 1)^2 =0,\nonumber\\
(\rho^{2}_{3/2} - 2\rho_{3/2} - 1)^2 =0,\nonumber\\
(\rho_{5/2} + 1)^2~(\rho^{2}_{5/2} - 4\rho_{5/2} + 1)^2 =0,\nonumber\\
(\rho^{4}_{7/2} - 4\rho^{3}_{7/2} - 6\rho^{2}_{7/2}+ 
4\rho_{7/2} +1)^2 =0,\nonumber\\
(\rho_{9/2} - 1)^2
(\rho^{4}_{9/2} - 4\rho^{3}_{9/2} - 14\rho^{2}_{9/2} - 4\rho_{9/2} + 1)^2 
=0,\nonumber\\
.~.~.~.~.~.~.~.~.~.~.~.~.~.~.~.~.~.~.~.~.~.~.~.~.~.~.~.~.~.~.~.~.~.~.~.~.~.\nonumber\\
\end{eqnarray}
and the eigenvalues of the density matrix are equal therefore to
$$
~~~~~~~~~~~~~~~~~~~~~~~~~1~~~~~~~~~~1~~~~~~~~~~~~~~~~~~~~~~~~~~~~~~~~~~~~~j=1/2        \nonumber
$$
$$
~~~~~~~~~~~~~~~~~1- \sqrt{2}~~~~~~~~\sqrt{2}+1~~~~~~~~~~~~~~~~~~~~~~~~~~~~~~~~j=3/2\nonumber
$$
$$
~~~~~~~~~~~~~~~~~2-\sqrt{3}~~~~-1~~~~2+\sqrt{3}~~~~~~~~~~~~~~~~~~~~~~~~~~~~~~~~j=5/2\nonumber
$$
$$
(1+\sqrt{2})+\sqrt{(\sqrt{2}+1)^2 +1}~~~~~~(1+\sqrt{2})-
\sqrt{(\sqrt{2}+1)^2 +1}~~~~~~~~~~~j=7/2\nonumber
$$
$$
(1-\sqrt{2})+\sqrt{(\sqrt{2}-1)^2 +1}~~~~~(1-\sqrt{2})-
\sqrt{(\sqrt{2}-1)^2 +1}~~~~~~~~~~~j=7/2\nonumber
$$
$$
1+\sqrt{5} + \sqrt{(\sqrt{5}+1)^2 -1}~~~~~~~~1+\sqrt{5} - 
\sqrt{(\sqrt{5}+1)^2 -1}~~~~~~~~~~~~j=9/2 \nonumber
$$
$$~~~~~~~~~~~~~~~~~~~~~~~~~~~~~~~~~~~~1~~~~~~~~~~~~~~~~~~~~~~~~~~~~~~~~~~~~~~~~~~~~~~~~~j=9/2\nonumber$$
$$
1-\sqrt{5} + \sqrt{(\sqrt{5}-1)^2 -1}~~~~~~~~1-\sqrt{5} -
\sqrt{(\sqrt{5}-1)^2 -1}~~~~~~~~~~~~j=9/2         \nonumber          
$$
\be
.~.~.~.~.~.~.~.~.~.~.~.~.~.~.~.~.~.~.~.~.~.~.~.~.~.~.~.~.~.~.~.~.~.~.~.~.~.~.~.~.~.~.~.~.\label{ospc}
\qee
Both states with $j=1/2$ have positive norms, the $j=3/2$ level has two 
positive and two negative norm states, the $j=5/2$ has four positive and two 
negative norm states, the $j=7/2$ has four positive and four negative norm states,
the $j=9/2$ have six positive and four negative norm states, and so on 
(see Fig.5). On the Figure 6 one can see the mass spectrum of the 
$\Sigma$-equation. The positive norm physical states, marked by the black 
bricks, are lying on the quasilinear trajectories and the negative norm ghost 
states, marked by the white bricks, are also lying on the quasilinear 
trajectories of different slope. Thus we have the equation which has the 
increasing 
mass spectrum, but the smallest mass on a given trajectory still tends to 
zero and in addition we have infinitely many ghost states. In the next section
we shall solve the problem of decreasing behaviour of the smallest mass on a 
given trajectory, but before that let us remark that one can 
project out the unwanted ghost states by constructing the projection
operator
\be
Pr = \prod_{i} (1+ \frac{1}{\epsilon_{i}} \Omega \Gamma_{0})
\qee
where the product is over ghost states. The details will be given in the second
part of this work where we shall formulate natural constraints
which appear in this system.

\begin{figure}
\centerline{\hbox{\psfig{figure=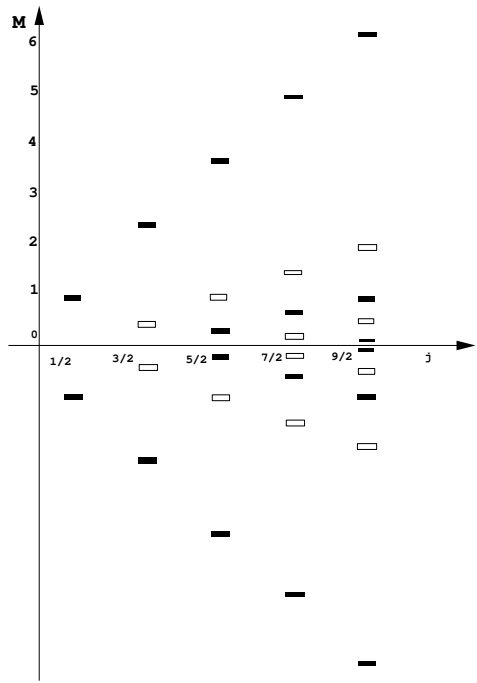,height=10cm,angle=0}}}
\caption[fig6]{The mass spectrum  of $H = M \Gamma^{-1}_{0}$. One can see 
positive norm physical states, marked by the black 
bricks, lying on the quasilinear trajectories and negative norm ghost 
states, marked by the white bricks, also lying on the quasilinear 
trajectories of different slope.}
\label{fig6}
\end{figure}

\subsection{Equation for the spin j components of the wave function}
Let us consider only a time dependent $\Sigma$-equation 
\be
i~\Gamma_{0}~\dot{\Psi} ~=~M~\Psi~~~~~~~i~\dot{\Psi} ~=\Gamma^{-1}_{0}~M~\Psi.
\qee
Using the explicit form of the gamma matrices (\ref{real}) one can get for 
$j=1/2$
\be
i~\left( \begin{array}{c}
  0,~~~1\\
  1,~~~0
\end{array} \right)  \partial_{t} \left( \begin{array}{c}
        \psi_{1}\\
        \psi_{\dot{1}}
\end{array} \right)_{1/2,m} = M~\left( \begin{array}{c}
        \psi_{1}\\
        \psi_{\dot{1}}
\end{array} \right)_{1/2,m}
\qee
and for $j=3/2$
\be
i~\left( \begin{array}{c}
  0,~~~1,~~~~~~~~~\\
  1,~~0,~~~2,~~~~\\~~~~~2,~~~0,~~~1\\~~~~~~~~~1,~~~0
\end{array} \right) \partial_{t} \left( \begin{array}{c}
        \psi_{2}\\
        \psi_{1}\\
        \psi_{\dot{1}}\\
        \psi_{\dot{2}}
\end{array} \right)_{3/2,m} = M~ \left( \begin{array}{c}
        \psi_{2}\\
        \psi_{1}\\
        \psi_{\dot{1}}\\
        \psi_{\dot{2}}
\end{array} \right)_{3/2,m}
\qee
and so on. After simple manipulations one can get equations for the $j=1/2$
components
\be
(\Box + M^{2}) \left( \begin{array}{c}
        \psi_{1}\\
        \psi_{\dot{1}}
\end{array} \right)_{1/2,m} =0,
\qee
and for $j=3/2$ components
\be
(\Box^{2} +6 M^{2}\Box + M^{4}) \left( \begin{array}{c}
        \psi_{2}\\
        \psi_{1}\\
        \psi_{\dot{1}}\\
        \psi_{\dot{2}}
\end{array} \right)_{3/2,m}=0
\qee
where $\Box =  \partial^{2}_{t}$
and for higher spins the differential operator will be of order $2j+1$ and it 
can be written by using the characteristic polynomial $p_{j}(\epsilon)$ of the 
$\gamma_{j}$ matrix
\be
p_{j}(\Box^{1/2})~\left( \begin{array}{c}
                     \psi_{j+1/2}\\
                     \psi_{j-1/2}\\
                      ........\\
                      ........\\
                      \psi_{\dot{j-1/2}}\\
                      \psi_{\dot{j+1/2}}
\end{array} \right)_{j,m}=0~~~~~~~~~m=-j,...,j
\qee
This Lorentz invariant operator has the order $2j+1$ and the first term is 
equal to the D'Alembertian in the power $j+1/2$ ~~,e.g. $\Box^{j+1/2}$ and  thus 
provides better convergence of quantum corrections.

\section{The solution with diagonal $\Gamma^{2}_{0}$}

In the case when some of the transition amplitudes in (\ref{realsym}) are set 
to zero 
\be
\gamma^{\dot{1}~1}~ = ~ \gamma^{2~3}_{j}~=~\gamma^{4~5}_{j} 
= ....=0~~~~~\gamma^{1~\dot{1}}~=~\gamma^
{\dot{2}~\dot{3}}~=~\gamma^{\dot{4}~\dot{5}}~= ... = 0 \label{diaggammsq}
\qee
and all other elements of the $\Gamma_{0}$ matrix remain the same as in 
(\ref{realsym}) we have a new $\Sigma_{1}$-solution with an important 
property that $\Gamma^{2}_{0}$ is a diagonal matrix and that the antihermitian 
part of $\Gamma_{k}$ anticommutes with $\Gamma_{0}$. Thus in
this case we recover the nondiagonal 
part of the Dirac commutation relations for gamma matrices
\be
\{ \Gamma_{0}, \tilde{\Gamma_{k}} \} =0~~~~~~~~k =x,y,z. \label{anticomm}
\qee
For the solution (\ref{diaggammsq}) 
one can explicitly compute the slope of the trajectories. For the first one 
we have
\be
M^{2}_{1} = 2M^{2}~(j-1) \approx 2M^{2} j,~~~~~~~~~~j=3/2,7/2,...
\qee
thus for this trajectory the string tension $\sigma$ is equal to 
\be
2\pi\sigma_{1}=\frac{1}{\alpha^{'}_{1}} = 2M^{2}.
\qee
For the second and third trajectories we have 
\be
M^{2}_{2} = M^{2} \frac{(j^2 - 3j +2)}{(j-1/2)} 
\approx M^{2} j,~~~~~~~~~~~j=5/2,9/2,...
\qee
\be
M^{2}_{3} = \frac{2M^{2}}{3} \frac{(j^2 - 5j +6)}{(j-1)} 
\approx \frac{2M^{2}}{3} j,~~~~~~~~~~~j=7/2,11/2,...
\qee
thus
\be
2\pi\sigma_{2}=\frac{1}{\alpha^{'}_{2}} = 
M^{2},~~~~~~~~~~~~~~~2\pi\sigma_{3}=\frac{1}{\alpha^{'}_{3}} = 2M^{2}/3,
\qee
and so on. When we move to the next trajectories the slope decreases 
and the string tension $\sigma_{n} = 1/2\pi\alpha^{'}_{n}$
varies from one trajectory to another and tends to zero
\be
2\pi \sigma_{n}  = \frac{1}{\alpha^{'}_{n}} = \frac{2M^2}{n} 
\rightarrow 0 \label{s}
\qee
{\it This result demonstrates that we have indeed the string equation which has 
trajectories with different string tension and that trajectories with large $n$
are almost "free" because the string tension tends to zero.}
Finally the general formula for all trajectories is
\be
M^{2}_{n}= \frac{2 M^{2}}{n}~ \frac{j^2 -(2n-1)j +n(n-1)}{j-(n-1)/2}~~n=1,2,...
\qee
where $j = n+1/2, n+5/2, ....$~(see Figure 7). The smallest mass   
on a given trajectory $n$ has spin $j = n+1/2$ and decreases as 
$$
M^{2}_{n}(j=n+1/2)= \frac{3M^2}{n(n+3)} .
$$
\begin{figure}
\centerline{\hbox{\psfig{figure=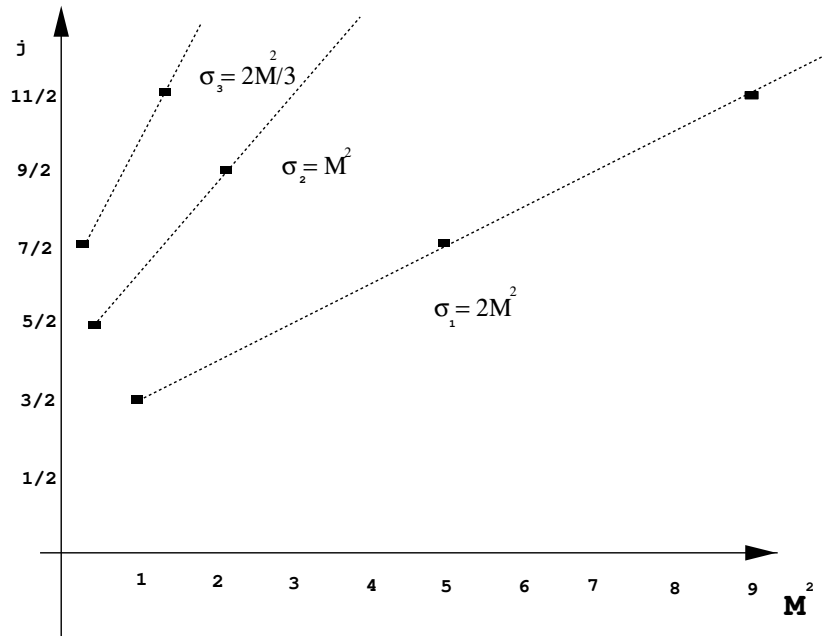,height=10cm,angle=0}}}
\caption[fig7]{The spectrum of $\Gamma^{-2}_{0}$ has a structure of 
linear trajectories with different string tension $2\pi\sigma_{n}=2M^{2}/n,~~n=
1,2,..$. The smallest mass on a given trajectory $n$
has the spin $j=n+1/2$ and tends to zero as $1/n^{2}$.}
\label{fig7}
\end{figure}

The other solution, $\Sigma_{2}$-solution, which shares the above properties 
of $\Sigma_{1}$-solution is (\ref{realsym}) with 
\be
\gamma^{1~2}_{j}~ = ~ \gamma^{3~4}_{j}~= ....
=0~~~~~~\gamma^{\dot{1}~\dot{2}}_{j}~=~\gamma^{\dot{3}~\dot{4}}_{j}~= ... = 0.
\label{delta2}
\qee
The difference between the last two solutions is that in the first case 
the lower spin is $j=3/2$ and in the second case it is $j=1/2$.
The unwanted property of all these solutions $\Sigma$, $\Sigma_{1}$ and 
$\Sigma_{2}$ is that the  smallest mass 
$M^{2}_{n}(min)$ tends to zero as it was in the original Majorana 
equation and the spectrum is not bounded from bellow. It is also true that 
$\Sigma = \Sigma_{1} + \Sigma_{2}$ in the sense of the footnote 5.

We have to remark also that both equations, $\Sigma_{1}$ and $\Sigma_{2}$, which 
correspond to  (\ref{diaggammsq}) 
and to (\ref{delta2}) have natural $constraints$. In the rest frame they are 
\be
\left( \begin{array}{c}
        \psi_{1}\\
        \psi_{\dot{1}}
\end{array} \right)_{1/2} = \left( \begin{array}{c}
        \psi_{3}\\
        \psi_{\dot{3}}
\end{array} \right)_{5/2}= .....=\left( \begin{array}{c}
        \psi_{2k-1}\\
        \psi_{\dot{2k-1}}
\end{array} \right)_{(4k-3)/2}~=~0~~~~k=1,2,...
\qee
for the $\Sigma_{1}$ case (\ref{diaggammsq}) and are 
\be
\left( \begin{array}{c}
        \psi_{2}\\
        \psi_{\dot{2}}
\end{array} \right)_{3/2} = \left( \begin{array}{c}
        \psi_{4}\\
        \psi_{\dot{4}}
\end{array} \right)_{7/2}= .....=\left( \begin{array}{c}
        \psi_{2k}\\
        \psi_{\dot{2k}}
\end{array} \right)_{(4k-1)/2}~=~0~~~~k=1,2,...
\qee
for the $\Sigma_{2}$ case (\ref{delta2}). We shall return to this 
constraints with more details in the second part of this article.

\section{ The $\Gamma_{5}$ mass term}

One can define the $\Gamma_{5}$ matrix which has all the properties of Dirac's 
$\gamma_{5}$ matrix (see Appendix C)~~~ $\{ \Gamma_{5},\Gamma_{\mu} \}=0$ and 
allows to introduce an additional mass term into the  
string equation (\ref{stringeq}) of the form   
$$  
(\vec{a}\cdot \vec{b}) ~ \Gamma_{5}.
$$
It was our aim to use the representations $\Theta$ of the Lorentz group 
which have nonzero Casimir operator 
$$
<j,m,r~\vert ~\vec{a}\cdot \vec{b}~ \vert~ r,j,m> =~ i~\lambda~ (r -1/2)
$$
$$
<j,m,\dot{r}~\vert ~\vec{a} \cdot \vec{b}~ \vert~ \dot{r},j,m>= -i~\lambda~(r -1/2).
$$
The matrix $(\vec{a}\cdot \vec{b}) ~ \Gamma_{5} $ is diagonal 
$$
<j,m,r~\vert (\vec{a}\cdot \vec{b}) ~ \Gamma_{5}
\vert ~r,j,m>~ =~ <j,m,\dot{r}~\vert (\vec{a}\cdot \vec{b}) ~ \Gamma_{5}
 \vert ~\dot{r},j,m> 
$$
$$
=i~\lambda~M~(-1)^{r+1}(r -1/2).
$$
With this new mass term the mass spectrum of the theory changes and we get 
trajectory 
$$
M^{2}_{1} = 2 M^{2}~(j-1)\cdot \lambda^{2} j(j+1) 
\approx 2(\lambda M)^{2} j^{3},
$$
thus for this trajectory
$$
2\pi\sigma_{1}= \frac{1}{\alpha^{'}_{1}} = 2 (\lambda M)^{2}\cdot j^{2}.
$$
For the second and third trajectories we have 
$$
M^{2}_{2} = M^{2} \frac{(j^2 - 3j +2)}{(j-1/2)} 
\cdot \lambda^{2} (j-1)j
\approx (\lambda M)^{2} j^{3},
$$
$$
M^{2}_{3} = \frac{2M^{2}}{3} \frac{(j^2 - 5j + 6)}{(j-1)} 
\cdot \lambda^{2} (j-2)(j-1)
\approx \frac{2(\lambda M)^{2}}{3} j^{3},
$$
thus
\be
2\pi\sigma_{2}=\frac{1}{\alpha^{'}_{2}} = 
(\lambda M)^{2} \cdot j^{2} ~~~~~~~~~~~~~2\pi\sigma_{3}=\frac{1}
{\alpha^{'}_{3}} = 2(\lambda M)^{2}/3 \cdot j^{2},
\qee
and so on. The appearance of $\lambda$ in these formulas together with $M$
demonstrates that this contribution to the spectrum is possible only if we 
use representations which have nonzero Casimir operators. The general 
formula for all trajectories is 

\be
M^{2}_{n}= \frac{2 (\lambda M)^{2}}{n} \frac{j^2 -(2n-1)j +n(n-1)}{j-(n-1)/2}
(j+1-n)(j+2-n)~~~~n=1,2,...
\qee
and the smallest mass on a given trajectory behaves as 

$$
M^{2}_{n}(j=n+1/2)=\frac{3(\lambda M)^2}
{n(n+3)}~\frac{3}{2}~\frac{5}{2}~~~~~~~j=n+1/2
$$
and still falls down as $1/n^{2}$.

\section{The dual string equation}

Under the dual transformation (\ref{dual}) 
$$
\Theta = (j_{0};\lambda)  \rightarrow (\lambda;j_{0})
= \Theta^{dual}
$$
the representation $\Theta$  (\ref{basicrep}) is transformed into its dual
\be
.....(\lambda; -5/2)~~(\lambda; -3/2)~~(\lambda; -1/2)~~(\lambda; 
1/2)~~(\lambda;3/2)~~(\lambda; 5/2)....      \label{dualrep}
\qee
and we shall consider here the case $\lambda = 1/2$ in order to have the Dirac
representation incorporated in $\Theta$. The solution which is dual to $\Sigma_{2}$
(\ref{realsym}) and  (\ref{delta2})  is equal to 
\be
\gamma^{r+1~r}_{j}=~\gamma^{r~r+1}_{j} =
\gamma^{\dot{r+1}~\dot{r}}_{j}= \gamma^{\dot{r}~\dot{r+1}}_{j} =
 \sqrt{(\frac{1}{4}-\frac{j^2 + j}{4r^2-1})}~~~~~r\geq j+3/2    \label{dualsym}
\qee
where $j=1/2,3/2,5/2,...$,~~$r=2,4,6,....$ and the rest of the elements are equal to zero
\be
\gamma^{\dot{1}~1}~ = ~ \gamma^{1~2}_{j}~=~\gamma^{3~4}_{j} 
= ....=0~~~~~\gamma^{1~\dot{1}
}~=~\gamma^{\dot{1}~\dot{2}}~=~\gamma^{\dot{3}~\dot{4}}~= ... = 0.
\qee
The Lorentz boost operators $\vec{b}$ are antihermitian in this case 
\be
b^{+}_{k} = -b_{k},
\qee
because the amplitudes $\varsigma$ (\ref{nond}) are pure imaginary and 
therefore the $\Gamma_{k}$ matrices are also antihermitian
\be
\Gamma^{+}_{k} = - \Gamma_{k}.
\qee
The matrix $\Omega$ changes and is now equal to the parity operator $P$, the
relation 
$$
\Omega~\Gamma^{+}_{\mu}= \Gamma_{\mu}~\Omega
$$
remains valid. The diagonal part of $\Gamma_{k}$ anticommutates
with $\Gamma_{0}$
as it was before (\ref{anticomm})
\be
\{ \Gamma_{0}, \tilde{\Gamma_{k}} \} =0~~~~~~~~k =x,y,z
\qee
\begin{figure}
\centerline{\hbox{\psfig{figure=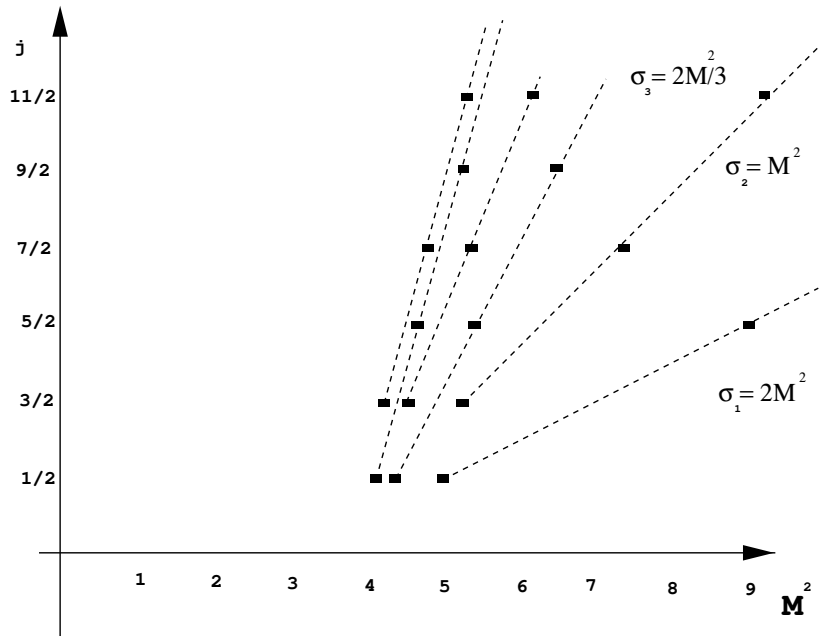,height=10cm,angle=0}}}
\caption[fig8]{The spectrum of the dual equation has again a structure of 
linear trajectories with  string tension 
$2\pi\sigma_{n}=2M^{2}/n,~~n=1,2,..$. The smallest mass on a given 
trajectory $n$ has the spin $j=1/2$ or $j=3/2$ and is bounded from bellow.}
\label{fig8}
\end{figure}
The mass spectrum is equal to
\be
M^{2}_{n}= \frac{2M^2}{n}~\frac{(j+n)(j+n+1)}{j+(n+1)/2} 
\label{dualmassspectrum}
\qee
where $n=1,2,3,..$ and enumerates the trajectories (see Figure 8). 
The lower spin on a given trajectory is either $1/2$ or $3/2$ 
depending on n: if n is odd then 
$j_{min}=1/2$, if n is even then $j_{min}=3/2$. This is an essential
new property
of the dual equation because now we have an infinite number of states 
with a given
spin $j$ instead of $j+1/2$, which we had before we did the dual transformation. 
The string tension is the same as in the dual system  (\ref{s})  
\be
 2\pi\sigma_{n}  =\frac{2M^2}{n}~~~~~~~~~~~~~~ n=1,2,3,...    \label{ss}
\qee
and the lower mass on a given trajectory $n$ is given by the formula $(j=1/2)$
\be
M^{2}_{n}(j=1/2) = \frac{4M^2}{n}\frac{(2n+1)(2n+3)}{n+2} \rightarrow (4M)^2.
\qee
{\it 
Thus the main problem of decreasing spectrum 
has been solved after the dual transformation because the spectrum is now bounded 
from below.}

Including the $\Gamma_{5}$ mass term one can see that all trajectories acquire a
nonzero slope
\be
M^{2}_{n}= \frac{2M^2}{n}~\frac{(j+n)^{2}(j+n+1)^{2}}{j+(n+1)/2} \frac{1}{4}
\qee
where
$$
n=1,2,3,....
$$
$$
j(n)_{min}=1/2,3/2,1/2,....
$$
thus for the lower mass on a given trajectory $n$ we have  
\be
M^{2}_{n}(j=1/2) = \frac{M^2}{16}\frac{(2n+1)^{2}(2n+3)^{2}}{n(n+2)} 
\rightarrow M^{2}n^{2}.
\qee
Turning on the pure Casimir mass term $gM~(\vec{a}^2 -\vec{b}^2)$ we will get the 
spectrum which grows as $j^{5}$
\be
M^{2}_{n}= g^{2}~\frac{2M^2}{n}\frac{(j+n)^{3}(j+n+1)^{3}}{j+(n+1)/2} 
\label{pureca},
\qee
where $n=1,2,3...$

\section{Extension to bosons}

Irrespective of the future interpretation of the theory it is important 
to include bosons.
The appropriate representation $\Theta$  for this purpose is 
\be
.....(\lambda; -3)~~(\lambda; -2)~~(\lambda; -1)~~(\lambda;0)~~~~~(\lambda;
0)~~(\lambda;1)~~(\lambda;2)~~(\lambda; 3)....      \label{}
\qee
and we shall consider $\lambda = 0$. The solution for $\Gamma_{0}$ is 
of the same form as we had before for fermions 
\be
\gamma^{r+1~r}_{j}=~\gamma^{r~r+1}_{j} =
\gamma^{\dot{r+1}~\dot{r}}_{j}= \gamma^{\dot{r}~\dot{r+1}}_{j} =
 \sqrt{( \frac{1}{4}-\frac{j^2 + j}{4r(r-1)} )}~~~~~r \geq j + 2    \label{boson}
\qee
where $j=0,1,2,3,..$,~~$r=2,4,6,....$ and the rest of the  elements 
are equal to zero
\be
\gamma^{\dot{1}~1}~ = ~ \gamma^{1~2}_{j}~=~\gamma^{3~4}_{j} 
= ....=0~~~~~\gamma^{1~\dot{1}
}~=~\gamma^{\dot{1}~\dot{2}}~=~\gamma^{\dot{3}~\dot{4}}~= ... = 0.
\qee
One can compute the slope of the first trajectory
\be
M^{2}_{1} = 2M^{2}~(j+2) \approx 2M^{2} j,~~~~~~~~~~j=0,2,4,...
\qee
thus for this trajectory the string tension $\sigma$ is equal to 
\be
2\pi\sigma_{1}=\frac{1}{\alpha^{'}_{1}} = 2M^{2}
\qee
and coincides with the one we obtain for the fermion string. The general formula
for the mass spectrum is 
\be
M^{2}_{n}= \frac{2M^2}{n}\frac{(j+n)(j+n+1)}{j+(n+1)/2}~~~~~n=1,2,3,...
\qee
and the lower spin on a given 
trajectory is either $0$ or $1$ depending on n: if n is odd then 
$j_{min}=0$, if n is even then $j_{min}=1$. Comparing this spectrum 
with the fermionic one (\ref{dualmassspectrum}) one can see that 
fermions and bosons lie on the same trajectories. In the bosonic case 
we do not have $\Gamma_{5}$-matrix and the Casimir operator 
$(\vec{a}\cdot\vec{b})$ is zero because $\lambda =0$, but we can include pure 
Casimir mass term $(\vec{a}^2 -\vec{b}^2)$ into the string equation to receive 
additional contribution to the slope of the trajectories. The formula for the 
spectrum is the same as for the fermions (\ref{pureca}).

\section{Discussion}
\vspace{.5cm}

In this article we suggest a relativistic equation for the gonihedric string
which has the Dirac form. The spectrum of the theory consists of 
particles and antiparticles of increasing half-integer spin 
lying on quasilinear trajectories of different slope. Explicit formulas for 
the mass spectrum allow to compute the string tension and thus demonstrate 
the string character of the theory. The old problem of decreasing mass 
spectrum 
has been solved and the spectrum is bounded from below. The trajectories are 
only asymptotically linear and thus are different from the free string 
linear trajectories. It is difficult to say at the moment what is the 
physical reason for this nonperturbative behaviour. Nonzero Casimir 
operators and the generalization of the $\Gamma_{5}$ matrix allow to introduce 
additional mass terms into the string equation and to increase the slope 
of the trajectories.

The equation is explicitly Lorentz invariant, but what we have now to worry about 
is the unitarity of the theory and unwanted ghost solutions.
In the second part of this article we shall return to these questions and will
demonstrate that tachyonic solutions which appear in Majorana equation do 
not show up here. This is because the nondiagonal transition amplitudes 
of the form
$<..j..\vert~\Gamma_{k}~\vert j \pm 1>$ are small here and the diagonal amplitudes 
$<..j..\vert~\Gamma_{k}~\vert j >$ are large. Indeed in the Majorana equation 
the diagonal elements are equal to zero, therefore the 
nondiagonal elements give rise 
to space-like tachyonic solutions (see (20) in \cite{majorana}). In the 
Dirac equation 
only diagonal transition amplitudes are present, and so 
the equation does not 
admit tachyonic solutions. The problem of ghost states is more subtle here 
and we shall analyze  
the natural constraints appearing in the system in the 
second part of this article to 
ensure that they decouple from the physical space of states.

In conclusion one of us (G.K.S.) wishes to acknowledge the hospitality of the 
Niels Bohr Institute 
where this work was started and would like to thank J.Ambjorn, P.Olesen, 
G.Tiktopoulos and P.Di Vecchia for the interesting discussions. 
We are thankful to Pavlos Savvidis, who has found the 
recurrent relation for the computation of the 
characteristic polynomials of the 
Jacoby matrices which finally leads to the basic solution for $\Gamma_{0}$.
This work has been 
initiated during Triangular Meeting in Rome, in March 1996, where we got the 
collected
articles of Ettore Majorana: "La vita e l'opera di Ettore Majorana (1906-1938)"
Roma, Accademia Nazionale dei Lincei, 1966, (ed. by E. Amaldi).

\begin{figure}
\centerline{\hbox{\psfig{figure=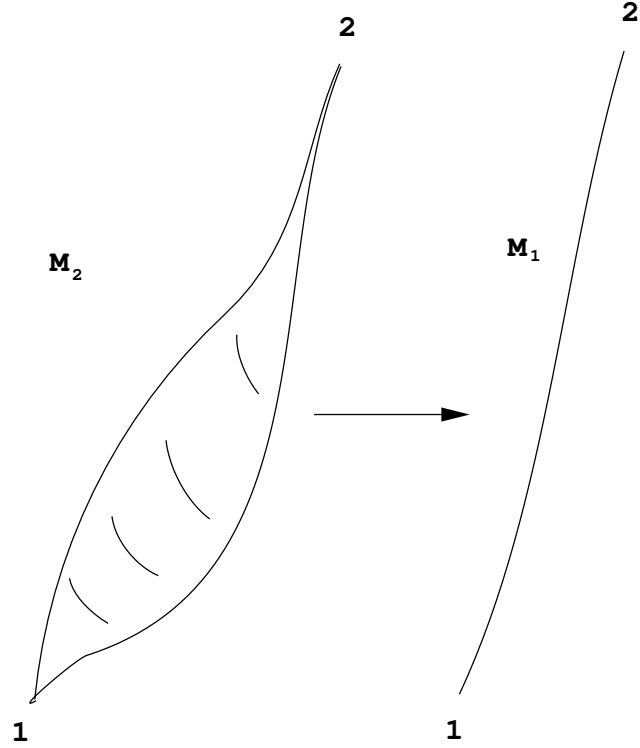,height=10cm,angle=0}}}
\caption[fig9]{Surface $M_{2}$ degenerates into a single world line $M_{1}$ 
and the action $A(M_{2})$ is reduced to the length of the world line $L(M_{1})$.}
\label{fig9}
\end{figure}

\section{Appendix A}
{\it Jacoby matrices.}~~ We are searching for the solution of equations 
(\ref{gamma}) in the form of Jacoby matrices 
\be
\left( \begin{array}{c}
  0,~~a\\
  a,~~0
\end{array} \right),~~ \left( \begin{array}{c}
  0,~~b,~~~~~~~\\b,~~0,~~a,~~~\\~~~~a,~~0,~~b\\~~~~~~~~b,~~0
\end{array} \right),~~ \left( \begin{array}{c}
  0,~~c,~~~~~~~~~~~~~~~\\
  c,~~0,~~b,~~~~~~~~~~~\\~~~~b,~~0,~~a,~~~~~~~\\~~~~~~~~a
,~~0,~~b,~~~\\~~~~~~~~~~~~b,~~0,~~c\\~~~~~~~~~~~~~~~~c
,~~~0
\end{array} \right),....\label{jacob}
\qee
The characteristic polynomials have the form
\begin{eqnarray}
x^2 - a^2 =0 \nonumber\\
x^4 -(a^2 + 2b^2) x^2 + b^4=0\nonumber\\
x^6 -(a^2 +2b^2 + 2c^2)x^4 +(2a^2 c^2 + 
2b^2 c^2 +b^4 +c^4)x^2 -a^2 c^4=0\nonumber\\
x^8 -(a^2 +2b^2 +2c^2 +2d^2)x^6 +~~~~\\
(2a^2 c^2 +2b^2 c^2 +2a^2 d^2 +2c^2 d^2 +
b^4 +c^4 +d^4 +4b^2 d^2)x^4-~~~~ \nonumber\\
-(a^2 c^4 +a^2 d^4 +2b^2 d^4 + 2b^4 d^2 +2a^ 2c^ 2d^2 
+2b^2 c^2 d^2)x^2 +b^4 d^4=0\nonumber  \\
.~.~.~.~.~.~.~.~.~.~.~.~.~.~.~.~.~.~.~.~.~.~.~.~.~.~.~.~.~.~.~.\nonumber\\
\end{eqnarray}
\section{Appendix B}
{\it $\sigma$ and $\tau$ algebra.}~~{{\bf j=1/2}}
\be
\sigma_{1/2}=0\nonumber
\qee
\vspace{.5cm}
{{\bf j=3/2}}
For this spin we have
\be
\sigma_{3/2},~~\sigma^{*}_{3/2},~~\sigma^{2}_
{3/2}~=~\sigma^{*~2}_{3/2}=0 \nonumber
\qee
and the main relation is
\be
\sigma_{3/2}~\sigma^{*}_{3/2} + \sigma^{*}_{3/2}~\sigma_{3/2}
= 4\nonumber
\qee
and thus
\begin{eqnarray}
\gamma_{3/2}= \omega_{3/2}(1+\sigma_{3/2})\\
\end{eqnarray}
\vspace{.5cm}
{{\bf j=5/2}}
For this spin we have
\be
\sigma_{5/2},~\sigma^{*}_{5/2},~\sigma^{2}_{5/2},~\sigma^{*~2}_{5/2}
,~~\sigma^{3}_{5/2}~=~\sigma^{*~3}_{5/2}~=0\nonumber
\qee
which appears to satisfy the algebra
\be
\{\sigma_{5/2},~\sigma^{*}_{5/2}\}
-2~(\sigma_{5/2} + \sigma_{5/2})
-3~\{\sigma^{2}_{5/2},~\sigma^{*~2}_{5/2}\}+
4~(\sigma^{2}_{5/2} + \sigma^{*~2}_{5/2})=2
\qee
Then it follows that
\begin{eqnarray}
\gamma_{5/2}= \omega_{5/2}(1+\sigma_{5/2})\nonumber\\
\end{eqnarray}
and that
\be
\sigma^{2}_{5/2}~\sigma^{*~2}_{5/2}~\sigma^{2}_{5/2}=
2~\sigma^{2}_{5/2}~\sigma^{*}_{5/2}~\sigma^{2}_{5/2}.
\qee
Similar algebra takes place for the $\tau_{j}$ matrices.

\section{Appendix C }
{\it Parity and $\Gamma_{5}$ matrices.}~~The commutation relations which 
define the parity operator P are
\be
P~a_{k} = a_{k}~P~~~~~P~b_{k} = - b_{k}~P~~~~~P^2 = 1,
\qee
because the parity operator P commutes with spatial rotations we have
\be
P = P^{rr'}_{j}~\delta_{jj'}~\delta_{mm'}
\qee
and from the second commutator 
\be
P^{r\dot{r}}_{j} = - P^{r\dot{r}}_{j-1}~~~~P^{\dot{r}r}_{j} = 
- P^{\dot{r}r}_{j-1},   
\qee
thus
\be
P^{r\dot{r}}_{j} = P^{\dot{r}r}_{j} = (-1)^{[j]}
\qee
One can check that
\be
P~\Gamma_{0} = \Gamma_{0}~P~~~~~P~\Gamma_{k} = - 
\Gamma_{k}~P~~~~~~~P~\Omega = \Omega~P
\qee
\be
P_{1/2} = \left( \begin{array}{c}
  0,~1\\
  1,~0
\end{array} \right),~~ P_{3/2} =  
\left( \begin{array}{c}~~~~~~~~~-1
\\  ~~~~~~-1~~\\  ~~~-1~~~~~\\  -1~~~~~~~~ \end{array} 
\right),~~P_{5/2} = 
\left( \begin{array}{c}~~~~~~~~~~~~1
\\  ~~~~~~~~~1~~\\  ~~~~~~1~~~~~\\  ~~~1~~~~~~~~\\~~1~~~~~~~~~~\\
1~~~~~~~~~~~~~ \end{array} \right),...
\qee
For $\Gamma_{5}$ we have 
\be
\Gamma_{5}~a_{k} = a_{k}~\Gamma_{5}~~~~~\Gamma_{5}~b_{k} = 
b_{k}~\Gamma_{5}~~~~~~~~~\Gamma_{5}^2 = 1
\qee
because it commutes with spatial rotations we have
\be
\Gamma_{5} = \Gamma_{5~j}^{rr'}~\delta_{jj'}~\delta_{mm'}
\qee
and from the second commutator
\be
\Gamma_{5~j}^{rr} =  \Gamma_{5~j-1}^{rr}~~~~\Gamma_{5~j}^{\dot{r}
\dot{r}} = \Gamma_{5~j-1}^{\dot{r}\dot{r}}~~~~\Gamma_{5~j}^{rr} 
= - \Gamma_{5~j}^{\dot{r}\dot{r}}
\qee
thus
\be
\Gamma_{5~j}^{rr} = -\Gamma_{5~j}^{\dot{r}\dot{r}} = (-1)^{r+1}
\qee
One can check that
\be
\Gamma_{5}~\Gamma_{0} = - 
\Gamma_{0}~\Gamma_{5}~~~~~\Gamma_{5}~\Gamma_{k} = -
\Gamma_{k}~\Gamma_{5}~~~~~~~\Gamma_{5}~P = 
- P~\Gamma_{5}~~~~~~\Gamma_
{5}~\Omega 
= - \Omega~\Gamma_{5}
\qee
and thus
\be
\Gamma_{5~1/2} = \left( \begin{array}{c}
  1~~\\
  ~~~-1
\end{array} \right),~~ \Gamma_{5~3/2}= \left( \begin{array}{c}
  -1,~~~~~~~\\~~~1,~~~~\\~~~~~~-1~~\\~~~~~~~~~1
\end{array} \right),~~\Gamma_{5~5/2}= \left( \begin{array}{c}
  1~~~~~~~~~~~~\\~~-1~~~~~~~~~\\~~~~~1~~~~~~~\\~~~~~~~-
1~~~~\\~~~~~~~~~~1~~\\~~~~~~~~~~~~-1
\end{array} \right),...
\qee

\section{Appendix D}
{\it ~Hermitian solution of $\Gamma_{0}$ for finite N.}~~The Hermitian 
solution of (\ref{gamma}) for the $\Gamma_{0}$ can be constructed by using our 
previous solution
\be
\gamma^{r+1~r}_{j}=-\gamma^{r~r+1}_{j} =
-\gamma^{\dot{r+1}~\dot{r}}_{j}= \gamma^{\dot{r}~\dot{r+1}}_{j} =
i~ \sqrt{(1-\frac{r^2}{N^2})~(\frac{j^2 + j}
{4r^2-1} -\frac{1}{4})}~~~~~j\geq r+1/2
\qee
\be
\gamma^{1~\dot{1}}_{j} = \gamma^{\dot{1}~1}_{j} = j+1/2
\qee
which for the low values of $j$ is:
$$
\gamma_{1/2} = \left( \begin{array}{c}
  0,~~~~~~1\\
  1,~~~~~~0
\end{array} \right),~~ \gamma_{3/2} = \left( \begin{array}{c}
  0,~~~~~~~~~~~~~i\sqrt{1-1/N^2},~~~~~~~~~~~~~~~~~~~~~~~\\
  -i\sqrt{1-1/N^2},~~~~~0,~~~~~~2
,~~~~~~~~~~~~~~~~~~~~~~~\\~~~~~~~~~~~~~~~~~~~~~~~~2
,~~~~~~0,~~~~~~~~~~i\sqrt{1-1/N^2}\\~~~~~~~~~~~~~~~~~~~~~~~~~~~~-i
\sqrt{1-1/N^2},~~~~~~~~~~~0
\end{array} \right),\nonumber
$$
$$
\gamma_{5/2} = \left( \begin{array}{c}
  0,~~~~~~i\sqrt{(1-4/N^2)(1/3)},~~~~~~~~~~~~~~~~~~~~~~\\
  -i\sqrt{(1-4/N^2)(1/3)},~~~~~0,~~~~~~~~~~i
\sqrt{(1-1/N^2)(8/3)},~~~~~~~~~~~~~~~~~\\~~~~~~~~~~~~~-i
\sqrt{(1-1/N^2)(8/3)},~~~~0,~~~~~~~~~~3
,~~~~~~~\\~~~~~~~~~~~~~~~~~~~~~~~~~~~~~~~~~~~~~~~~~3
,~~~~~~~~~~0,~~~~~i\sqrt{(1-1/N^2)(8/3)}
,~~~~~~~~~\\~~~~~~~~~~~~~~~~~~~~~~~~~~-i\sqrt{(1-1/N^2)(8/3)},~~~0
,~~~i\sqrt{(1-4/N^2)
(1/3)}\\~~~~~~~~~~~~~~~~~~~~~~~~~~~~~~~~~~~~~~~-i
\sqrt{(1-4/N^2)(1/3)},~~~0
\end{array} \right),\nonumber
$$
$$etc...$$
The characteristic equations for these matrices are:
\begin{eqnarray}
\gamma^{2}_{1/2}-1=0,\nonumber\\
(\gamma^{2}_{3/2}-2\gamma_{3/2}-1+1/N^2)~(
\gamma^{2}_{3/2}+2\gamma_{3/2}-1+1/N^2)=0,\nonumber\\
(\gamma^{3}_{5/2}+3\gamma^{2}_{5/2}-3\gamma_{5/2}-1+ 
4(\gamma_{5/2}+1)/N^2)~~~~\nonumber\\
(\gamma^{3}_{5/2}-3\gamma^{2}_{5/2}-
3\gamma_{5/2}+1+
4(\gamma_{5/2}-1)/N^2) =0,\nonumber\\
............................................................\nonumber
\end{eqnarray}
and the eigenvalues $\epsilon_{j}$ can be found
$$
        1~~~~~~~~~~~~~~~~~~~~~~~~~~~~~j=1/2\nonumber
$$
$$
\sqrt{2-1/N^2}-1~~~~~~~~\sqrt{2-1/N^2}
+1~~~~~~~~~~~~~~~~~~~~~~~~~~~~j=3/2\nonumber
$$
$$
.~.~.~.~.~.~.~.~.~.~.~.~.~.~.~.~.~.~.~.~.~.~.~.~.~.~.~.~.~.~.~.~.~.~.~.~.~.~.~.~.~.~.~.~.~
~~~~~~~~~~~~~\nonumber
$$
When $N \rightarrow \infty$ these formulas are reduced to the 
formulas presented in 
the section where we consider the Hermitian solution.

\section{Appendix E}
{\it Nonzero Casimir operators. }~~If $\lambda =0$ then 
$$ 
<\vec{a}\cdot\vec{b}> = 0~~~<(\vec{a}^2 -\vec{b}^2)> = (r - 1/2)^2 -1\nonumber
$$
and 
$$
\varsigma^{r}_{j} = \varsigma^{\dot{r}}_{j}= 
\frac{1}{2} \sqrt{1-(\frac{r^2 -r}{j^2 -1/4})}~~~~~~j \geq r+1/2. \nonumber
$$
and if $\lambda = 1$ then 
$$
<\vec{a}\cdot\vec{b}> = i~(r - 1/2)~~~~~<(\vec{a}^2 -\vec{b}^2)> 
= (r - 1/2)^2 \nonumber
$$ 
An essential simplification appears when $\lambda =1/2$, in that case 
$$
\varsigma^{r}_{j} = 
\varsigma^{\dot{r}}_{j}= 
\frac{1}{2} \sqrt{1-(\frac{r-1/2}{j})^2}~~~~~~j \geq r+1/2\nonumber
$$
and 
$$
<\vec{a}\cdot\vec{b}> = i/2~(r - 1/2)~~~~~<(\vec{a}^2 -\vec{b}^2)> 
= (r - 1/2)^2 - 3/4.  \nonumber
$$

\vfill
\end{document}